\documentclass[11pt]{article}

\usepackage[utf8]{inputenc}
\usepackage[T1]{fontenc}
\usepackage{lmodern}
\usepackage{microtype}

\usepackage{amsmath,amssymb,amsfonts}
\usepackage{graphicx}
\usepackage{subcaption}
\usepackage{booktabs}
\usepackage{multirow}
\usepackage{tabularx}
\usepackage{array}
\usepackage{enumitem}
\usepackage{xcolor}
\usepackage{colortbl}
\usepackage{listings}
\usepackage[skins,breakable]{tcolorbox}
\usepackage{float}
\usepackage{tikz}
\usetikzlibrary{arrows.meta,positioning,shapes.geometric,fit,backgrounds,calc}

\usepackage[
    colorlinks=true,
    linkcolor=blue!60!black,
    citecolor=blue!60!black,
    urlcolor=blue!60!black
]{hyperref}

\usepackage[nameinlink,noabbrev]{cleveref}
\usepackage[margin=1in]{geometry}

\definecolor{rqblue}{RGB}{30,100,160}
\definecolor{rqbluebg}{RGB}{235,245,255}
\definecolor{ctxorange}{RGB}{160,80,10}
\definecolor{ctxorangebg}{RGB}{255,245,230}
\definecolor{epdgviolet}{RGB}{100,40,160}
\definecolor{epdgvioletbg}{RGB}{245,235,255}
\definecolor{headergray}{RGB}{220,228,238}
\definecolor{bestgreen}{RGB}{230,250,235}
\definecolor{tealtrack}{RGB}{0,120,100}
\definecolor{violettrack}{RGB}{100,40,160}

\newtcolorbox{rqfinding}[1]{
  enhanced,
  breakable,
  colback=rqbluebg,
  colframe=rqblue,
  fonttitle=\bfseries\small\sffamily,
  coltitle=white,
  attach boxed title to top left={yshift=-2mm, xshift=4mm},
  boxed title style={colback=rqblue, arc=2pt, boxrule=0pt},
  title={#1},
  boxrule=0.7pt,
  arc=3pt,
  left=6pt, right=6pt, top=7pt, bottom=5pt,
}

\newtcolorbox{ctxfinding}[1]{
  enhanced,
  breakable,
  colback=ctxorangebg,
  colframe=ctxorange,
  fonttitle=\bfseries\small\sffamily,
  coltitle=white,
  attach boxed title to top left={yshift=-2mm, xshift=4mm},
  boxed title style={colback=ctxorange, arc=2pt, boxrule=0pt},
  title={#1},
  boxrule=0.7pt,
  arc=3pt,
  left=6pt, right=6pt, top=7pt, bottom=5pt,
}

\newtcolorbox{epdgfinding}[1]{
  enhanced,
  breakable,
  colback=epdgvioletbg,
  colframe=epdgviolet,
  fonttitle=\bfseries\small\sffamily,
  coltitle=white,
  attach boxed title to top left={yshift=-2mm, xshift=4mm},
  boxed title style={colback=epdgviolet, arc=2pt, boxrule=0pt},
  title={#1},
  boxrule=0.7pt,
  arc=3pt,
  left=6pt, right=6pt, top=7pt, bottom=5pt,
}

\lstdefinestyle{prompttemplate}{
    backgroundcolor=\color{gray!4},
    rulecolor=\color{gray!50},
    commentstyle=\color{green},
    keywordstyle=\color{magenta},
    numberstyle=\tiny\color{gray},
    stringstyle=\color{purple},
    basicstyle=\ttfamily\tiny,
    breakatwhitespace=false,         
    breaklines=true,                 
    captionpos=b,                    
    keepspaces=true,                 
    numbers=left,                    
    numbersep=5pt,                  
    showspaces=false,                
    showstringspaces=false,
    showtabs=false,                  
    tabsize=2
}

\newcommand{\tool}{\textsc{RepBench}}

\newcommand{\hdr}[1]{\textbf{#1}}
\newcolumntype{R}[1]{>{\raggedleft\arraybackslash}p{#1}}

\title{
\LARGE\bfseries Representation Matters:\\
\Large An Empirical Study of
Program 
Representations for LLM Vulnerability Reasoning
}

\author{
Andrew Stoltman 
\quad
Johnathan Tang
\quad
Haipeng Cai \\
Department of Computer Science and Engineering\\
University at Buffalo, SUNY \\
\texttt{\{aestoltm,jtang,haipengc\}@buffalo.edu}
}

\date{}

\begin{document}

\maketitle

\begin{abstract}
Large Language Models (LLMs) are increasingly adopted for automated vulnerability detection, but the optimal way to represent program structure and semantics for LLM-based vulnerability reasoning remains unclear. Most existing prompting-based approaches expose the model directly to raw source code, implicitly assuming that more source-level context provides better evidence. This paper challenges that assumption through a systematic empirical benchmark of raw source code and various static-analysis-based code representations for LLM vulnerability analysis.

We present \tool{}, a unified benchmark pipeline that converts real-world C/C++ vulnerability testcases into multiple program representations---raw source code, Abstract Syntax Trees (ASTs), Control-Flow Graphs (CFGs), Program Dependence Graphs (PDGs), and representation combinations thereof---along with an auxiliary track of enriched Program Dependence Graphs (ePDGs). Using a curated PrimeVul-derived corpus of 107 standard Joern-based testcases across five CWE categories, we evaluate ten representation variants under a fixed Chain-of-Thought (CoT) prompting and structured-output protocol, with an additional 19 auxiliary ePDG cases generated through an external VulChecker/Hector pipeline. 

Our results show that representation choice has a substantial effect on LLM vulnerability reasoning. The strongest variant, AST+PDG, achieves 83.2\% curated 
accuracy compared with 53.5\% for the raw-source baseline---a 29.7 percentage point improvement. At the prompt-family level, graph-only prompts outperform both source-only and source-plus-graph prompts while requiring substantially lower prompt overhead. This reveals a counter-intuitive \emph{context dilution} effect: adding raw source code to compact structural graph evidence often degrades reasoning performance rather than improving it. Our findings suggest that carefully selected structural representations offer a stronger accuracy--prompt-overhead tradeoff than simply exposing LLMs to more raw input, and that static analysis can serve as an effective prompt-construction layer for security-focused LLM reasoning.
\end{abstract}

\section{Introduction}
\label{sec:intro}

LLMs have rapidly become a major component of modern software engineering workflows, including code generation, debugging, test generation, program repair, and vulnerability analysis. In software security, LLMs are increasingly used to inspect code, explain suspicious behaviors, identify vulnerable statements, and reason about whether a function is vulnerable or safe~\cite{sheng2025llmssoftwaresecuritysurvey,zhou2024largelanguagemodelvulnerability}. Despite this growing adoption, a foundational design question remains underexplored: \emph{what form of program representation should be given to an LLM when asking it to reason about security-relevant code behavior?}

Most existing LLM-based vulnerability analysis pipelines use raw source code as the primary input representation~\cite{sheng2025llmssoftwaresecuritysurvey}. This is natural because LLMs are trained on large corpora of source code and can interpret syntax, idioms, and API usage directly. However, vulnerability reasoning often depends on program relations that are not explicit in surface syntax. Identifying an integer overflow may require understanding whether arithmetic operations are guarded by appropriate bounds checks~\cite{cwe_mitre}. Detecting a double free or use-after-free may require reasoning about allocation, deallocation, pointer aliases, and subsequent reuse~\cite{cwe_mitre}. These properties are precisely what software-analysis representations---e.g., control-flow graphs and program dependence graphs---have been designed to expose.

This paper studies the hypothesis that \emph{structured program representations can help LLMs reason about vulnerability semantics more effectively than raw source code, because they expose security-relevant structural evidence in a compact and focused form}. We evaluate this hypothesis through a controlled representation-ablation benchmark. The vulnerability detection task is held fixed: for each testcase, the LLM must classify a target function as \texttt{VULNERABLE} or \texttt{SAFE} for a specified CWE. The only experimental factor varied is the input representation provided to the model.

Accordingly, we build \tool{}, a benchmark framework that transforms real-world vulnerability testcases into multiple textualized program representations and evaluates them under a fixed LLM vulnerability-detection task. The current study draws from PrimeVul-derived C/C++ examples~\cite{ding2024vulnerabilitydetectioncodelanguage} and focuses on five CWE categories: CWE-122, CWE-190, CWE-191, CWE-415, and CWE-416. The main evaluation track uses Joern~\cite{joern_paper} to extract AST, CFG, and PDG views; an auxiliary track evaluates enriched PDGs (ePDGs) generated through the VulChecker/Hector pipeline~\cite{mirskyvulchecker}. Across all variants, the task remains the same: given a target function and a specified CWE, the model must classify the function as \textsc{Vulnerable} or \textsc{Safe}. Only the program representation given to the model is varied.

Specifically, our study is organized around four main research questions:

\begin{itemize}
\item \textbf{RQ1:}
How effectively do standalone structural graph representations support LLM-based vulnerability reasoning compared to raw source code?
\item \textbf{RQ2:}
Which program representations and representation combinations are most effective for LLM vulnerability reasoning? 
\item \textbf{RQ3:}
What tradeoff exists between representation effectiveness and prompt overhead?
\item \textbf{RQ4:} How do representation effects vary across vulnerability categories, ground-truth labels, and auxiliary ePDG inputs?
\end{itemize}

Our evaluation reveals a strong and counter-intuitive representation effect on LLM performance in vulnerability reasoning, including four main findings aligned with these questions. 

For \textbf{RQ1}, \textit{structural graph representations substantially outperform raw source-code prompting}. The raw-source baseline achieves only 53.5\% curated 
accuracy, while every evaluated graph-based variant improves over it. At the prompt-family level, graph-only prompts reach 74.4\% accuracy.

For \textbf{RQ2}, representation choice matters not only at the family level but also among individual structural views, and \textit{combining complementary graph views is especially effective}. The strongest variant is AST+PDG, which achieves 83.2\% curated 
accuracy, a 29.7 percentage-point improvement over raw source. AST+CFG is the second strongest variant, achieving 80.4\% curated accuracy. These results suggest that LLM vulnerability reasoning benefits from combining local syntactic anchors with either dependence evidence or control-flow evidence. However, source-plus-graph variants do not consistently improve over, and often underperform, graph-only variants, indicating that adding raw source code can introduce distracting context---not useful complementary evidence.

For \textbf{RQ3}, graph-only representations provide the best effectiveness--overhead tradeoff. They achieve the highest family-level
accuracy while using substantially fewer prompt characters and input tokens than source-plus-graph prompts. In particular, source-plus-graph prompts require 2.4$\times$ as many characters and 2.2$\times$ as many input tokens as graph-only prompts, yet achieve lower
accuracy. This result reveals what we call the \emph{context dilution} effect: adding more input context can reduce the salience of vulnerability-relevant evidence, causing the model to attend to irrelevant or misleading content despite receiving more total information. Our case studies further show that this effect is not merely a consequence of prompt truncation; source-augmented prompts can fail even when they remain well within the prompt budget.

For \textbf{RQ4}, representation effects are broadly visible across the adequately supported CWE categories, but they are not uniform. AST+PDG improves over raw source on CWE-190, CWE-415, and CWE-416, suggesting that \textit{the aggregate graph advantage is not solely an artifact of the CWE-190-heavy corpus}. At the same time, source-plus-graph prompts perform competitively on some memory-lifecycle cases, indicating that raw source context may still help for certain vulnerability patterns. Label-wise analysis further shows that the model is more reliable on SAFE cases than on VULNERABLE cases. Finally, the auxiliary ePDG track shows that \textit{vulnerability-oriented dependence metadata can support correct reasoning in some cases}, but its current effectiveness is limited by LLVM-level readability and artifact-generation feasibility.

Together, these findings have implications beyond this specific benchmark. LLM-based software analysis systems that rely solely on raw source prompting may fail to exploit decades of progress in static analysis. Conversely, \textit{representation-aware systems can use program analysis as a semantic compression layer}, presenting models with compact structural evidence rather than unfiltered source code. This paper provides an initial empirical foundation for designing such representation-aware LLM vulnerability-analysis systems.

In summary, this paper makes the following contributions in LLM-based vulnerability analysis. 

\begin{itemize}[leftmargin=*]
    \item We present \tool{}, a benchmark framework for evaluating how various (raw source code and static-analysis-based) program representations affect LLM vulnerability reasoning, including a PrimeVul-derived evaluation corpus of 107 standard Joern-track testcases across five CWE categories and a separate 19-case auxiliary ePDG track (\Cref{sec:repbench}).
    \item We systematically evaluate raw source code, ASTs, CFGs, PDGs, graph combinations, source-plus-graph hybrids, and ePDG prompts under a fixed LLM vulnerability detection task (\Cref{sec:setup}).
    \item We identify and characterize a context dilution effect: graph-only prompts outperform larger source-augmented prompts even when truncation is absent, demonstrating that more input context does not necessarily improve LLM vulnerability reasoning. 
    We show that compact structural representations provide a stronger accuracy--prompt-overhead tradeoff than raw source-code prompting, motivating the use of static analysis as a prompt-construction layer for LLM-based security reasoning (\Cref{sec:results}).
\end{itemize}

\section{Background and Motivation}

\subsection{LLM-Based Vulnerability Analysis}

LLM-based vulnerability analysis has attracted substantial interest because LLMs can interpret code, natural-language vulnerability descriptions, and security rules within a unified prompting interface. Recent work has explored LLMs for vulnerability detection, explanation, localization, and repair~\cite{zhou2024largelanguagemodelvulnerability}. Many such systems ask the model to inspect source code and determine whether it contains a given vulnerability class. Techniques include fine-tuning on labeled vulnerability corpora, prompt engineering with chain-of-thought guidance, and retrieval-augmented prompting~\cite{liu2023lostmiddlelanguagemodels,nong2025appatch,li2025svtrustevalcevaluatingstructuresemantic}.

Despite this progress, the representation question has received comparatively little attention. Most pipelines assume that presenting source code directly to the model is sufficient and that the model will internally reconstruct the program relations needed for vulnerability reasoning. This assumption may not hold in practice: vulnerabilities often depend on semantic relations that are scattered across statements, branches, and data dependencies, and the evidence for them may be easily overlooked in a long, unstructured source-code prompt.

\subsection{Static-Analysis-Based Representations}

This study focuses on four families of program representations, each of which exposes a different structural view of the target function.

\paragraph{Abstract Syntax Trees.}
ASTs capture the hierarchical syntactic structure of a function~\cite{sunabstract26tosem}. They preserve operators, expressions, statements, nested control constructs, calls, declarations, and lexical structure. For vulnerability reasoning, ASTs expose syntactic patterns such as arithmetic expressions, pointer dereferences, memory-management calls, and validation checks---the local anchors from which security-relevant reasoning must begin. However, ASTs do not encode execution order or data propagation, so they cannot alone reveal whether a guard actually dominates a dangerous operation.

\paragraph{Control-Flow Graphs.}
CFGs capture possible execution orders among program statements or basic blocks~\cite{aho2006compilers}. They are important for reasoning about whether a guard condition dominates a dangerous operation and whether validation occurs before use. For arithmetic vulnerabilities, a CFG can reveal whether a bounds check occurs before an overflow-prone computation. For memory-lifecycle bugs, CFG evidence can help distinguish mutually exclusive cleanup paths from paths that may execute repeated frees. CFGs encode reachability and execution order, but they do not directly capture how values or pointer states propagate between statements---that information requires dependence analysis.

\paragraph{Program Dependence Graphs.}
PDGs combine data-dependence and control-dependence relations~\cite{pdg_paper}. They are particularly relevant for vulnerability analysis because many security bugs depend on relationships between sources, guards, sinks, and state updates that are spread across the function body. A PDG can make explicit that a function parameter flows into an arithmetic expression, that an allocation result later reaches a free operation, or that a pointer-state update controls a subsequent use~\cite{pdg_paper}. This makes PDGs especially valuable for source-to-sink reasoning in memory-lifecycle vulnerabilities. We note that Joern-derived PDGs expose relevant data and control dependence edges, but they should not be interpreted as resolving all pointer aliases or fully characterizing heap-object lifetimes; their value lies in surfacing dependency evidence that would require significant reasoning to reconstruct from source text alone.

\paragraph{Enriched Program Dependence Graphs.}
The auxiliary ePDG track uses the VulChecker/Hector toolchain~\cite{mirskyvulchecker}. Compared with standard Joern-derived PDGs, ePDG artifacts include vulnerability-oriented metadata such as operation types, semantic tags, node data types, conditions, link types, and dependency categories. These artifacts are generated from LLVM intermediate representation (IR) using a custom LLVM plugin that annotates candidate root-cause and manifestation locations~\cite{mirskyvulchecker}. ePDGs may provide richer vulnerability evidence, but they require external build feasibility and project-specific artifact generation, making them substantially less reliable across real-world projects.

\begin{figure}[tp]
\centering
\begin{tikzpicture}[
  >=Stealth, thick,
  font=\small\sffamily,
  inode/.style={draw, rectangle, text centered, minimum height=0.68cm,
                inner xsep=5pt, inner ysep=3pt, text width=#1,
                fill=orange!14, draw=orange!60, line width=0.8pt},
  onode/.style={draw, rectangle, text centered, minimum height=0.68cm,
                inner xsep=5pt, inner ysep=3pt, text width=#1,
                fill=green!12, draw=green!55!black, line width=0.8pt},
  pnode/.style={draw, rounded corners=4pt, text centered, minimum height=0.65cm,
                inner xsep=5pt, inner ysep=3pt, text width=#1,
                fill=blue!6, draw=blue!45, line width=0.7pt},
  jnode/.style={draw, rounded corners=4pt, text centered, minimum height=0.65cm,
                inner xsep=4pt, inner ysep=3pt, text width=4.55cm,
                fill=teal!10, draw=teal!55, line width=0.7pt},
  llmnode/.style={draw, rounded corners=-2pt, text centered, minimum height=0.85cm,
                inner xsep=12pt, inner ysep=5pt, text width=#1,
                fill=black!10, draw=black!55, line width=3.2pt},
  enode/.style={draw, rounded corners=4pt, text centered, minimum height=0.65cm,
                inner xsep=4pt, inner ysep=3pt, text width=4.55cm,
                fill=violet!10, draw=violet!55, line width=0.7pt},
  arr/.style={->, thick},
  jarr/.style={->, thick, teal!65!black},
  earr/.style={->, thick, violet!65!black},
  metaarr/.style={->, thick, dashed, orange!65!black},
  stlabel/.style={font=\scriptsize\bfseries\sffamily, text=gray!55!black,
                  rotate=90, align=center},
]
 
\node[stlabel] at (-7.6,-1.1)  {Stage 1: Inputs};
 
\node[inode=8.5cm] (pv)  at (0, 0)   {\textbf{PrimeVul Dataset}\quad (C/C++, CVE/CWE metadata)};
\node[pnode=10.5cm] (cur) at (0,-1.1) {Testcase Curation \& Metadata Verification};
\node[pnode=10.5cm] (chk) at (0,-2.2) {Source Repository Checkout};
 
\node[inode=2.2cm] (meta) at (7.8,-1.1)
  {Testcase\\Metadata\\({\footnotesize CWE, CVE,\\source location})};
 
\draw[gray!85, thin, dashed] (-8.0,-2.68) -- (8.3,-2.68);
 
\node[stlabel] at (-7.4,-4.5) {Stage 2: Code\\Representation\\Generation};
 
\node[font=\scriptsize\bfseries\sffamily, text=teal!70!black]   at (-2.15,-3.07) {Standard Joern Track};
\node[font=\scriptsize\bfseries\sffamily, text=violet!70!black] at ( 2.15,-3.07) {Auxiliary ePDG Track};
 
\node[jnode] (joern)  at (-4.15,-3.75) {Joern CPG Extractor};
\node[onode=4.0cm] (grph)   at (-4.15,-4.80) {AST / CFG / PDG};
\node[jnode] (jser)   at (-4.15,-5.85) {Graph Serializer};
 
\node[enode] (vc)     at ( 4.15,-3.75) {VulChecker/Hector Pipeline};
\node[onode=4.0cm] (ejsonl) at ( 4.15,-4.80) {ePDG Records (JSONL)};
\node[enode] (eser)   at ( 4.15,-5.85) {ePDG Serializer};
 
\draw[gray!85, thin, dashed] (-8.0,-6.38) -- (8.3,-6.38);
 
\node[stlabel] at (-7.6,-8.55) {Stage 3: Outputs};
 
\node[pnode=10.5cm] (pc) at (0,-7.10)
  {\textbf{Prompt Constructor} (50{,}000-char budget $\cdot$ CWE guidance $\cdot$ few-shot)};
\node[llmnode=10.5cm] (llm) at (0,-8.20)
  {\textbf{Large Language Model} (\texttt{qwen3.6-plus} $\cdot$ $T{=}0.2$ $\cdot$ CoT)};
\node[pnode=10.5cm] (parser) at (0,-9.30)
  {\textbf{Output Parser \& Evaluator} (effective \& curated accuracy)};
\node[onode=8.5cm] (result) at (0,-10.35)
  {\textbf{Classification Result:} \texttt{VULNERABLE} / \texttt{SAFE}};
 
 
\draw[arr] (pv)  -- (cur);
\draw[arr] (cur) -- (chk);
 
\draw[metaarr] (cur.east) -- (meta.west);
\draw[metaarr] (meta.south) |- (pc.east);
 
\draw[jarr] (chk.south) -- ++(0,-0.25) -| (joern.north);
\draw[earr] (chk.south) -- ++(0,-0.25) -| (vc.north);
 
\draw[jarr] (joern) -- (grph);
\draw[jarr] (grph)  -- (jser);
 
\draw[earr] (vc)     -- (ejsonl);
\draw[earr] (ejsonl) -- (eser);
 
\draw[jarr] (jser.south) -- ++(0,-0.28) -| ([xshift=2.6cm]pc.north west);
\draw[earr] (eser.south) -- ++(0,-0.28) -| ([xshift=-2.6cm]pc.north east);
 
\draw[arr] (pc)     -- (llm);
\draw[arr] (llm)    -- (parser);
\draw[arr] (parser) -- (result);
 
\end{tikzpicture}
\caption{Overview of the \tool{} benchmark pipeline across three stages.
\textbf{Stage~1 (Inputs):} PrimeVul vulnerability testcases are curated and checked out from source repositories; testcase metadata (CWE type, CVE identifier, source location) is extracted and passed directly to the prompt constructor (dashed arrow).
\textbf{Stage~2 (Code Representation Generation):} Source code is processed through two parallel tracks---the standard Joern track (teal, left) extracts AST, CFG, and PDG views; the auxiliary ePDG track (violet, right) processes VulChecker/Hector artifacts---and each track's output is serialized into compact text.
\textbf{Stage~3 (Outputs):} Serialized representations and testcase metadata are combined by the prompt constructor, the LLM produces a structured reasoning trace, and the output parser evaluates the binary classification result.
Input/output data nodes are shown as sharp rectangles; 
processing steps use rounded rectangles.}
\label{fig:overview}
\vspace{-6pt}
\end{figure}
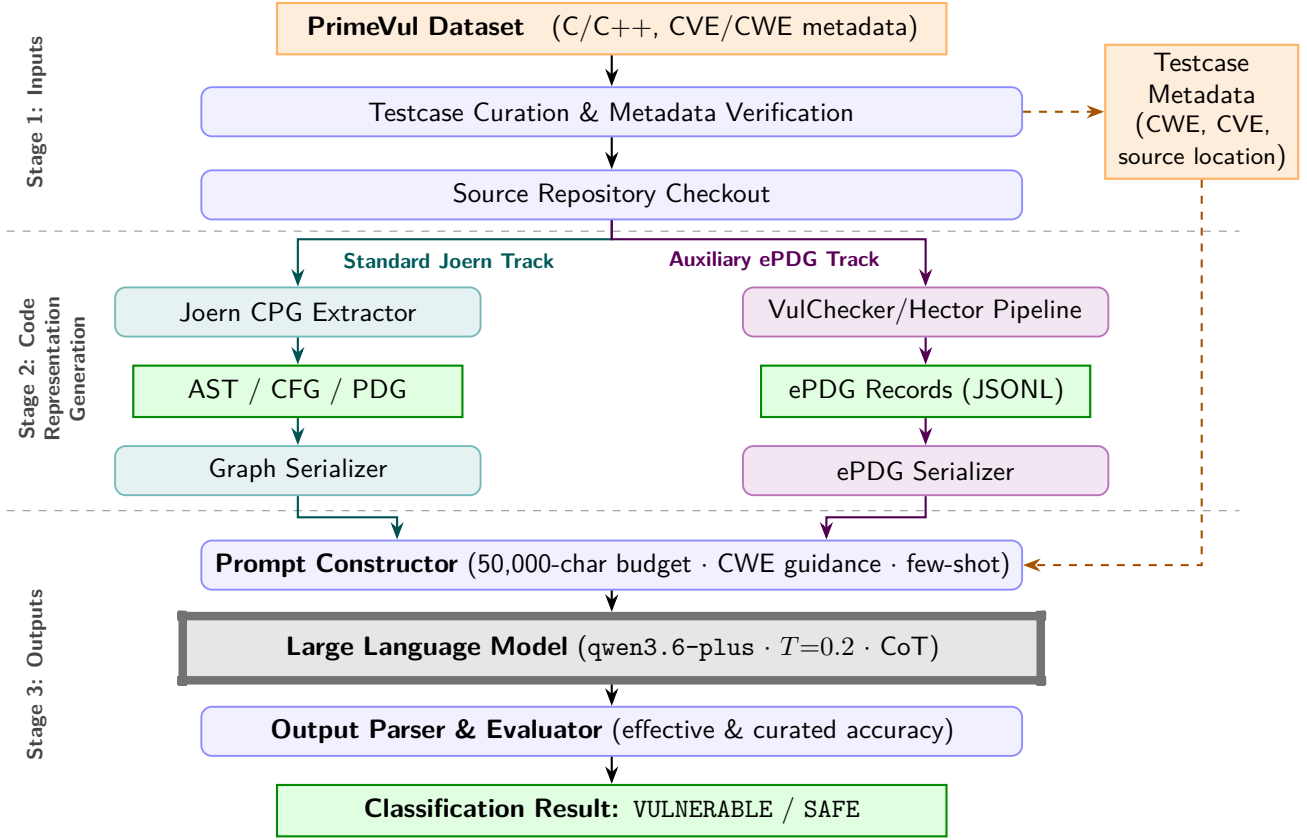

\subsection{Why Representation Matters for LLM Reasoning}

Representation design is especially consequential for LLM-based software analysis because prompts are constrained by context length, inference cost, and the model's capacity for sustained attention~\cite{liu2023lostmiddlelanguagemodels}. Raw source code may contain declarations, formatting artifacts, comments, macros, helper logic, and statements unrelated to the vulnerability under examination. In contrast, structural graphs can act as a semantic compression layer: they discard some surface detail while making control and data relations explicit.

This creates a key tradeoff. A representation must preserve enough security-relevant semantics to support vulnerability reasoning, yet remain compact enough to fit within realistic prompt budgets. If a prompt grows too long or too noisy, the model may fail to attend to the relevant evidence~\cite{liu2023lostmiddlelanguagemodels}---the \emph{context dilution} effect described in \Cref{sec:intro}. Our benchmark is designed specifically to measure this tradeoff across a diverse set of representation choices.

\vspace{-3pt}
\section{Design of {\tool}}\label{sec:repbench}
\vspace{-3pt}
We start with an overview of our benchmark framework. Then, we elaborate how we build the evaluation dataset, generate various program representations, and construct different families of prompts using those program representations. 

\vspace{-6pt}
\subsection{Benchmark Overview}\label{sec:bench}

\vspace{-2pt}
\subsubsection{Design Goals}
\vspace{-2pt}
The benchmark is designed around three principles as per the goal of our study. \textbf{First}, the vulnerability detection task must remain fixed while the input representation changes. This allows performance differences to be attributed to representation choice rather than task reformulation. \textbf{Second}, structural program evidence must be serialized into compact textual forms suitable for LLM prompting: raw formats such as DOT or JSONL contain substantial markup that is useful for tools but inefficient for reasoning. \textbf{Third}, the pipeline must preserve metadata for reproducibility, including testcase identifiers, source locations, representation variants, prompt sizes, clipping status, model configuration, and output labels.

\subsection{Pipeline Architecture}

\tool{} is organized as an end-to-end pipeline with six stages: dataset selection and verification, testcase normalization, program representation construction, representation parsing and serialization, prompt generation, and LLM-based vulnerability analysis (\Cref{fig:overview}). Each testcase begins as a function-level vulnerability example with project metadata, CWE/CVE information, source-location fields, and a binary vulnerability label. The benchmark verifies testcase metadata, loads the checked-out source file, constructs the requested representations, serializes them into prompt-compatible text, and queries the model to classify the target function.

The pipeline contains two representation-generation branches. The main Joern-based branch extracts AST, CFG, and PDG representations through the Joern static analysis platform. The auxiliary VulChecker/Hector branch produces ePDG artifacts when build and artifact-generation requirements are satisfied. The central unit of analysis is the target function: the benchmark imports the full source file into Joern to improve parsing stability, but exports graphs focused on the target method to reduce irrelevant context and prompt overhead.

\subsection{Dataset Construction}

\subsubsection{PrimeVul-Derived Corpus}

The benchmark draws on PrimeVul as its primary source of real-world vulnerability data~\cite{ding2024vulnerabilitydetectioncodelanguage}. PrimeVul provides 6,968 vulnerable and 228,800 fixed function-level C/C++ examples extracted from open-source projects, paired with vulnerability metadata including CWE and CVE fields across 140 CWE categories. Its function-level granularity aligns well with the target-method scope of the benchmark, and its paired vulnerable/fixed structure makes it straightforward to derive balanced binary-classification testcases.

The benchmark uses paired vulnerable and fixed function examples from PrimeVul. Each retained testcase has five components: project and vulnerability metadata, the checked-out source file, target-function boundary information, one or more derived program representations, and evaluation records produced from prompts constructed over those representations. Testcase generation also clones the corresponding project repository when ePDG artifacts require a buildable project context.

\paragraph{Corpus Filtering.}
The 107-case standard corpus was derived from the full PrimeVul set through sequential filtering. First, CWE types outside the five supported categories (CWE-122, -190, -191, -415, -416) were excluded. Second, testcases with missing or malformed CWE/CVE metadata were removed. Third, testcases for which source repository checkout failed were excluded. Fourth, testcases for which Joern parsing or graph extraction produced empty or structurally invalid outputs were removed. Fifth, duplicate or ambiguous testcase entries were excluded. The resulting 107-case corpus represents the intersection of semantic scope and extraction feasibility under the current pipeline. The selected CWE scope therefore reflects both security relevance and practical extraction feasibility; future work should expand the pipeline to support a broader CWE range.

The current benchmark focuses on C/C++ function-level testcases. These languages provide a rich set of memory and arithmetic vulnerabilities aligned with the selected CWE categories, and Joern provides mature support for C/C++ parsing and analysis. Nevertheless, our study findings should be interpreted as evidence for C/C++ vulnerability reasoning rather than as language-independent conclusions.

\subsubsection{Supported CWE Categories}

The five supported CWE categories span two complementary vulnerability classes:

\begin{itemize}[leftmargin=*]
    \item \textbf{CWE-122}: Heap-based Buffer Overflow---memory write past allocated bounds.
    \item \textbf{CWE-190}: Integer Overflow---signed or unsigned wraparound in arithmetic.
    \item \textbf{CWE-191}: Integer Underflow---arithmetic result below representable minimum.
    \item \textbf{CWE-415}: Double Free---deallocation of already-freed memory.
    \item \textbf{CWE-416}: Use After Free---access to deallocated memory.
\end{itemize}

Arithmetic vulnerabilities (CWE-190, CWE-191, CWE-122) require reasoning about value ranges, guard conditions, and the relationship between bounds checks and potentially unsafe operations~\cite{cwe_mitre}. Memory-lifecycle vulnerabilities (CWE-415, CWE-416) require reasoning about allocation, deallocation, pointer state, and reuse across the function body~\cite{cwe_mitre}. This mix allows the benchmark to enable testing whether different graph representations support different styles of security-relevant reasoning.

\subsubsection{Standard Joern Track and Auxiliary ePDG Track}

The benchmark maintains two distinct evaluation tracks. The \emph{standard Joern track} includes 107 unique PrimeVul-derived testcases from 30 real-world projects. These testcases support Joern-based extraction of AST, CFG, and PDG representations and constitute the primary representation-ablation study.

The \emph{auxiliary ePDG track} contains 19 retained ePDG cases. ePDG generation depends on additional build feasibility constraints and external project-specific artifact availability, making it infeasible for the majority of PrimeVul testcases. Within the 19 retained cases, an 11-case overlapping subset supports direct paired comparison with the Joern-based variants. 

These two tracks should be interpreted as related but distinct evaluations; aggregate accuracy figures from the ePDG track are not directly comparable to the 107-case Joern results except on the overlapping subset.

\subsubsection{Corpus Statistics}

\begin{table}[t]
\centering
\caption{Standard Joern-track corpus composition.}
\label{tab:corpus}
\setlength{\tabcolsep}{10pt}
\begin{tabular}{lr}
\toprule
\rowcolor{headergray}
\hdr{Category} & \hdr{Count} \\
\midrule
Unique testcases  & 107 \\
Distinct projects & 30  \\
CWE types         & 5   \\
SAFE testcases    & 52  \\
VULNERABLE testcases & 55 \\
\bottomrule
\end{tabular}
\end{table}

\begin{table}[t]
\centering
\caption{Standard corpus distribution by CWE category.}
\label{tab:cwe}
\setlength{\tabcolsep}{8pt}
\begin{tabular}{llr}
\toprule
\rowcolor{headergray}
\hdr{CWE} & \hdr{Description} & \hdr{Testcases} \\
\midrule
CWE-122 & Heap-based Buffer Overflow & 4  \\
CWE-190 & Integer Overflow           & 56 \\
CWE-191 & Integer Underflow          & 2  \\
CWE-415 & Double Free                & 23 \\
CWE-416 & Use After Free             & 22 \\
\midrule
\multicolumn{2}{l}{Total}            & 107 \\
\bottomrule
\end{tabular}
\end{table}

\Cref{tab:corpus} shows that the standard Joern corpus is nearly balanced overall (52 SAFE, 55 VULNERABLE). However, \Cref{tab:cwe} reveals a non-uniform per-CWE distribution: CWE-190 accounts for more than half of all testcases, while CWE-122 and CWE-191 contain only a handful of cases. Aggregate results should therefore be interpreted as corpus-level trends rather than equally strong conclusions for every individual CWE category.

\subsection{Program Representation Construction}

\subsubsection{Joern-Based Graph Extraction}

The main representation branch uses Joern, a static analysis platform that parses C/C++ code into a Code Property Graph (CPG)~\cite{joern_paper}. From this unified representation, the benchmark extracts three structural views of each target function: AST, CFG, and PDG. The goal is not merely to test whether graphs in general help, but to examine which structural views provide the best tradeoff between vulnerability evidence and prompt overhead.

\subsubsection{Extraction Scope}

The main extraction strategy is denoted \texttt{full\_file\_target\_method}. The complete checked-out source file is first imported into Joern, and analysis is executed over the full file; the exported graph is then restricted to the target method. This balances parsing fidelity---the full file preserves surrounding declarations, macros, and type context---against prompt focus, since restricting the export to the target method prevents irrelevant file-level context from inflating the prompt. An auxiliary \texttt{full\_file\_all\_methods} scope is implemented for future ablation but is not the mainline configuration.

\subsubsection{Representation Rationale}
\label{sec:reprationale}

Each representation is included because it exposes a distinct view of the evidence relevant to vulnerability reasoning. The goal is to isolate how different structural abstractions affect LLM reasoning ability under a fixed task and prompting protocol.

\paragraph{Raw source code.}
Raw source code is the natural baseline because most prompting-based LLM vulnerability analysis systems present code directly to the model~\cite{sheng2025llmssoftwaresecuritysurvey}. It preserves the full lexical and syntactic context available to a human reviewer, including declarations, expressions, comments, local helper logic, and surrounding implementation detail. This baseline tests whether the LLM can reconstruct the vulnerability semantics directly from source text. At the same time, real-world source files frequently contain macros, helper routines, formatting noise, and code unrelated to the target vulnerability---all of which can inflate the prompt with context that dilutes the security-relevant evidence.

\paragraph{Abstract Syntax Trees.}
ASTs capture local syntactic structure, exposing operators, call expressions, variable declarations, statement nesting, pointer dereferences, casts, and memory-management calls in a form that is less dependent on surface formatting than raw code. Many CWE patterns have recognizable local syntactic anchors: arithmetic expressions for CWE-190 and CWE-191, allocation and deallocation calls for CWE-415 and CWE-416, and buffer-access operations for CWE-122. ASTs, however, do not encode execution order or data propagation; they are tested to assess how far local structural syntax can support classification without explicit flow or dependence information.

\paragraph{Control-Flow Graphs.}
CFGs represent possible execution orders among program operations. Many vulnerabilities depend not only on whether a dangerous operation exists, but also on whether it is guarded by validation on all relevant execution paths. For arithmetic vulnerabilities, a CFG can reveal whether a bounds check dominates an overflow-prone computation. For memory-lifecycle vulnerabilities, control-flow evidence can distinguish mutually exclusive cleanup paths from paths that may execute repeated frees. CFGs expose reachability and guard structure, but they do not capture how values or pointer states propagate between statements---a limitation that motivates their combination with PDGs.

\paragraph{Program Dependence Graphs.}
PDGs combine data-dependence and control-dependence relations~\cite{pdg_paper}. They are especially relevant for vulnerability reasoning because security bugs often depend on relationships between sources, guards, and sinks that are separated across the function. A PDG can make explicit that a parameter flows into an arithmetic expression, that an allocation result reaches a subsequent free, or that a pointer-state update controls a later use. This directly supports source-to-sink reasoning for memory-lifecycle bugs. PDGs may, however, be less self-contained than source or AST inputs because dependency edges can omit local syntactic detail needed to interpret the precise operation at each node.

\paragraph{Enriched Program Dependence Graphs.}
The auxiliary ePDG representation explores whether richer, vulnerability-oriented dependence metadata can further assist LLM reasoning. Unlike Joern PDGs, ePDG records may include operation categories, node data types, conditions, semantic tags, link types, and link data types---additional cues about guards, arithmetic operations, memory operations, and candidate root-cause or manifestation sites. ePDG generation is build-dependent and substantially less reliable across real-world projects; it is therefore treated as an auxiliary track rather than part of the primary ablation.

\paragraph{Representation combinations.}
Graph combinations are included because different representations capture complementary evidence. \texttt{ast\_pdg} combines local syntactic anchors with cross-statement dependency information, which is useful when a vulnerability depends on both the exact operation at a site and the flow of values or pointer state reaching that site. \texttt{ast\_cfg} combines local syntax with execution-order and guard information. \texttt{cfg\_pdg} combines path structure with dependency evidence. Source-plus-graph variants (\texttt{full}, \texttt{ast\_plus\_source}, \texttt{pdg\_plus\_source}) test a different hypothesis: whether graphs are more useful as standalone substitutes for raw source or as augmentations to it. Together, these variants allow the benchmark to separate representation complementarity from the context dilution introduced by larger prompts.

\subsubsection{Representation Variants}

\Cref{tab:variants} summarizes the ten Joern-based prompt variants. These variants support direct comparison of raw source, individual graphs, graph-only combinations, and source-plus-graph hybrids, addressing 
supporting the main representation and prompt-overhead research questions. 

\begin{table}[t]
\centering
\caption{Joern-based prompt variants in the standard representation-ablation track.}
\label{tab:variants}
\setlength{\tabcolsep}{5pt}
\begin{tabular}{lccccl}
\toprule
\rowcolor{headergray}
\hdr{Variant} & \hdr{Source} & \hdr{AST} & \hdr{CFG} & \hdr{PDG} & \hdr{Purpose} \\
\midrule
\texttt{raw}             & \checkmark & --         & --         & --         & Source-only baseline \\
\texttt{ast}             & --         & \checkmark & --         & --         & Syntax-only graph \\
\texttt{cfg}             & --         & --         & \checkmark & --         & Control-flow-only graph \\
\texttt{pdg}             & --         & --         & --         & \checkmark & Dependence-only graph \\
\texttt{ast\_cfg}        & --         & \checkmark & \checkmark & --         & Syntax + control flow \\
\texttt{ast\_pdg}        & --         & \checkmark & --         & \checkmark & Syntax + dependence \\
\texttt{cfg\_pdg}        & --         & --         & \checkmark & \checkmark & Control flow + dependence \\
\texttt{full}            & \checkmark & \checkmark & \checkmark & \checkmark & Maximum-information hybrid \\
\texttt{ast\_plus\_source} & \checkmark & \checkmark & --       & --         & Source augmented with AST \\
\texttt{pdg\_plus\_source} & \checkmark & --       & --         & \checkmark & Source augmented with PDG \\
\bottomrule
\end{tabular}
\end{table}

\subsubsection{Auxiliary ePDG Construction}

The auxiliary ePDG branch processes retained VulChecker/Hector artifacts and converts them into prompt-compatible text. Unlike the Joern branch, ePDG generation depends on build feasibility and project-specific artifact availability; ePDG evaluation is therefore reported separately from the 107-case standard corpus.

ePDG records preserve granular semantic fields: operation, node data type, condition, semantic tag, link type, and link data type. To maintain approximate comparability with the Joern \texttt{full\_file\_target\_method} setting, the benchmark prioritizes \texttt{function\_only} ePDG files when available, falling back to full-file ePDG artifacts as a secondary condition.

\subsection{Representation Serialization and Prompt Construction}

\subsubsection{Graph Serialization}

Raw graph outputs are not directly suitable for LLM prompting. Joern exports DOT graphs, while the ePDG branch produces JSONL edge records. Both formats contain visualization markup, structural boilerplate, and tool-specific syntax that inflate prompt length without aiding reasoning. \tool{} therefore includes a dedicated parsing and serialization layer that converts graph artifacts into compact textual representations before prompt construction.

For Joern-generated graphs, a DOT conversion script parses node identifiers, node types, code labels, source line numbers, and labeled edges. The converter linearizes the graph into a line-centric textual format grouped by source line, appending compact dependency summaries. This keeps the serialized representation close to program order while preserving selected structural links. Although edge categories (AST, CFG, DDG, CDG) are not printed explicitly in the current format, the variants differ in \emph{which} graph files are serialized: the \texttt{ast} variant feeds only the Joern AST output, \texttt{pdg} feeds only the PDG output, and combinations feed the corresponding graph files. The structural information distinguishing variants thus arises from which graph is serialized, not from per-edge type annotations in the output text.

For ePDG inputs, a JSONL conversion interface extracts edge and node metadata from retained VulChecker/Hector artifacts. The serialized ePDG text begins with compact statistics and risk-focused summaries, then lists prioritized edge records including operation names, node data types, conditions, semantic tags, link types, and link data types. To control prompt growth, the reported experiments retain at most 450 prioritized edges per testcase.

\subsubsection{Node and Edge Encoding}

The serialized Joern representation uses a lightweight line-oriented format. For each graph, the converter prints a short header, lists retained nodes grouped by source line, and appends a dependency summary when labeled edges are available. A typical node row has the form \texttt{L \textless{}line\textgreater{} | \textless{}node type\textgreater{} | \textless{}code label\textgreater{}}. For example, one row may record a \texttt{calloc} call at a given line, while another records a guard comparison or return statement.

Nodes without source-line information are omitted from the listing to reduce noise. When labeled dependency edges are present, the converter emits a \texttt{Dependencies:} block containing up to 15 summaries. These entries preserve Joern's internal node identifiers as lightweight cross-references; the security-relevant information is conveyed through line numbers, node types, code labels, and dependency labels. Graph elements are ordered by increasing source-line number, keeping the serialized form close to program order.

ePDG serialization uses a different edge-centric format. Each retained record is printed as \texttt{edge\_id: source\_node\,--LINK:dtype-->\,target\_node}, preceded by compact summary statistics. In the reported auxiliary experiments, the converter retains at most 450 prioritized ePDG edges per testcase.

\subsubsection{Serialization Example}
\label{sec:serexample}

As a concrete illustration, consider a FreeRDP~\cite{FreeRDP} function (\texttt{msusb\_mspipes\_read}) that validates stream capacity before allocating an array of pipe descriptors. The source-level evidence is:

\begin{lstlisting}
if (Stream_GetRemainingCapacity(s) / 12 < NumberOfPipes)
    return NULL;
MsPipes = calloc(NumberOfPipes, sizeof(...));
\end{lstlisting}

The DOT graph exported by Joern is not inserted into the prompt directly. Instead, the serializer emits compact line-oriented text:

\begin{lstlisting}
L 67 | <operator>.division    | Stream_GetRemainingCapacity(s) / 12
L 67 | <operator>.lessThan    | Stream_GetRemainingCapacity(s) / 12 < NumberOfPipes
L 68 | return                 | return NULL
L 70 | calloc                 | calloc(NumberOfPipes, sizeof(...))
Dependencies:
  node_42 -> node_67 [NumberOfPipes]
  node_42 -> node_70 [NumberOfPipes]
  node_67 -> node_70 [Stream_GetRemainingCapacity(s) / 12 < NumberOfPipes]
\end{lstlisting}

This reflects the actual prompt-level structure: the model sees source-line entries, node types, code labels, and a dependency block rather than raw DOT visualization syntax. The dependency entries use Joern internal identifiers as cross-references, and security-relevant information is conveyed through line numbers, node types, and dependency labels. In this case, the PDG makes explicit that \texttt{NumberOfPipes} flows into both the guard at line~67 and the allocation at line~70, and that the guard controls execution to the allocation---evidence directly relevant to whether the CWE-190 check is valid.

\subsubsection{LLM Prompts Used}

The benchmark uses three prompt-template families that differ in which program evidence is included. Source-only contains \texttt{raw}; graph-only contains \texttt{ast}, \texttt{cfg}, \texttt{pdg}, \texttt{ast\_cfg}, \texttt{ast\_pdg}, and \texttt{cfg\_pdg}; and source-plus-graph contains \texttt{full}, \texttt{ast\_plus\_source}, and \texttt{pdg\_plus\_source}. \Cref{tab:prompttemplates} summarizes the common instruction blocks and the representation-specific content without reproducing the long prompts in full.

\begin{table}[t]
  \centering
  \caption{Compact structure of the three prompt-template families. Bracketed text denotes content inserted for each testcase.}
  \label{tab:prompttemplates}
  \small
  \setlength{\tabcolsep}{4pt}
  \begin{tabularx}{\linewidth}{>{\raggedright\arraybackslash}p{0.15\linewidth}XXX}
    \toprule
    \rowcolor{headergray}
    \hdr{Block} & \hdr{Source-only} & \hdr{Graph-only} & \hdr{Source+Graph} \\
    \midrule
    Task & Analyze C/C++ code for [CWE]. & Analyze a program representation for [CWE]. & Analyze C/C++ code and program representations for [CWE]. \\
    \midrule
    CWE guidance & \multicolumn{3}{l}{CWE-specific sink hints; identify sink, trace source, check validation, and analyze flow.} \\
    \midrule
    Representation & \texttt{[SOURCE CODE]} & \texttt{[AST/CFG/PDG TEXT]} & \texttt{[SOURCE CODE]} plus \texttt{[GRAPH TEXT]} \\
    \midrule
    Examples & \multicolumn{3}{l}{Dynamically selected CWE-matched few-shot examples that fit the prompt budget.} \\
    \midrule
    Output & \multicolumn{3}{l}{JSON: reasoning, \texttt{VULNERABLE}/\texttt{SAFE}, confidence, CWE mapping, and explanation.} \\
    \bottomrule
  \end{tabularx}
\end{table}

\paragraph{Source-only prompts.}
The source-only template (\Cref{lst:source_only}) 
presents the raw target source code inside a C/C++ code block and asks the LLM to determine whether the function contains the specified CWE. This serves as the primary baseline, corresponding to the common workflow of prompting a model directly with code.

\begin{lstlisting}[style=prompttemplate, language=Python, caption={Source-only prompt template}, label={lst:source_only}]
"""Analyze the following C/C++ code for potential issues related to [CWE ID] [CWE NAME].

## Task
Review the code to identify if issues exist for [CWE ID] [CWE NAME]. Focus on the target CWE only.

## Input Code
[SOURCE CODE]


## Few-Shot Examples
Example [EXAMPLE #] [LABEL]

[EXAMPLE SOURCE CODE]

Reasoning:
[EXAMPLE COT REASONING]
Label: [LABEL]

## Chain-of-Thought Checklist
Use the following reasoning steps internally and summarize the evidence in JSON:
1. Identify Sink: find dangerous operations for [CWE ID]. Look for: [CWE HINTS].
2. Trace Source: [TRACE STEP].
3. Check Validation: look for bounds checks, null checks, size checks, or pointer state updates before the sink.
4. Analyze Flow: [FLOW STEP].
5. Determine: classify as VULNERABLE or SAFE and assign confidence.

## Output Format
Return only valid JSON with this schema:
{{
  "reasoning": {{
    "sink_identified": "line and operation",
    "data_source": "origin of dangerous data or pointer",
    "validation_present": true,
    "control_flow_analysis": "whether validation guards the sink",
    "pdg_dependencies": ["relevant DDG/CDG evidence"]
  }},
  "conclusion": "VULNERABLE or SAFE",
  "confidence": 0.0,
  "cwe_mapping": "<CWE ID>",
  "explanation": "brief explanation"
}}"""
\end{lstlisting}
\vspace{-4pt}
\begin{lstlisting}[style=prompttemplate, language=Python, caption={Graph-only prompt template}, label={lst:graph_only}]
"""Analyze the following program representation for potential issues related to [CWE ID] [CWE NAME].

## Task
Review the graph text to identify if issues exist for [CWE ID] [CWE NAME]. Focus on the target CWE only. Code snippets are embedded inside graph node labels.

## Program Representations
[GRAPH REPRESENTATIONS]

## Few-Shot Examples
Example [EXAMPLE #] [LABEL]

[EXAMPLE SOURCE CODE]

Reasoning:
[EXAMPLE COT REASONING]
Label: [LABEL]

## Chain-of-Thought Checklist
Use the following reasoning steps internally and summarize the evidence in JSON:
1. Identify Sink: find dangerous operations for [CWE ID]. Look for: [CWE HINTS].
2. Trace Source: [TRACE STEP].
3. Check Validation: look for bounds checks, null checks, size checks, or pointer state updates before the sink.
4. Analyze Flow: [FLOW STEP].
5. Determine: classify as VULNERABLE or SAFE and assign confidence.

## Output Format
Return only valid JSON with this schema:
{{
  "reasoning": {{
    "sink_identified": "line and operation",
    "data_source": "origin of dangerous data or pointer",
    "validation_present": true,
    "control_flow_analysis": "whether validation guards the sink",
    "pdg_dependencies": ["relevant DDG/CDG evidence"]
  }},
  "conclusion": "VULNERABLE or SAFE",
  "confidence": 0.0,
  "cwe_mapping": "<CWE ID>",
  "explanation": "brief explanation"
}}"""
\end{lstlisting}

\paragraph{Graph-only prompts.}
The graph-only template (\Cref{lst:graph_only}) presents serialized AST, CFG, PDG, or combination evidence without raw source code. The prompt tells the model that code snippets are embedded inside graph node labels and that it should analyze the structural text as program evidence. This avoids instructing the model to inspect raw code that is not present.

\paragraph{Source-plus-graph prompts.}
The source-plus-graph template (\Cref{lst:Source_plus_graph_only}) presents both raw source code and one or more serialized graph views, framing the graph text as structural evidence that complements the source. These variants test whether graph representations are more useful as standalone substitutes for raw source or as augmentations to it.
\begin{lstlisting}[style=prompttemplate, language=Python, caption={Source-plus-graph prompt template}, label={lst:Source_plus_graph_only}]
"""Analyze the following C/C++ code and program representations for potential issues related to [CWE ID] [CWE NAME].

## Task
Review the code structure and data flows to identify if issues exist for [CWE ID] [CWE NAME]. Focus on the target CWE only.

## Input Code
[SOURCE CODE]

## Program Representations
AST lists code structure by source line. CFG dependencies describe execution order. PDG dependencies describe data/control flow; DDG means data dependence and CDG means control dependence.

[GRAPH REPRESENTATIONS]

## Few-Shot Examples
Example [EXAMPLE #] [LABEL]

[EXAMPLE SOURCE CODE]

Reasoning:
[EXAMPLE COT REASONING]
Label: [LABEL]

## Chain-of-Thought Checklist
Use the following reasoning steps internally and summarize the evidence in JSON:
1. Identify Sink: find dangerous operations for [CWE ID]. Look for: [CWE HINTS].
2. Trace Source: [TRACE STEP].
3. Check Validation: look for bounds checks, null checks, size checks, or pointer state updates before the sink.
4. Analyze Flow: [FLOW STEP].
5. Determine: classify as VULNERABLE or SAFE and assign confidence.

## Output Format
Return only valid JSON with this schema:
{{
  "reasoning": {{
    "sink_identified": "line and operation",
    "data_source": "origin of dangerous data or pointer",
    "validation_present": true,
    "control_flow_analysis": "whether validation guards the sink",
    "pdg_dependencies": ["relevant DDG/CDG evidence"]
  }},
  "conclusion": "VULNERABLE or SAFE",
  "confidence": 0.0,
  "cwe_mapping": "<CWE ID>",
  "explanation": "brief explanation"
}}"""
\end{lstlisting}

\subsubsection{CWE-Aware Reasoning Guidance}

Each prompt includes CWE-aware reasoning guidance. The prompt inserts a CWE name and a set of CWE-specific sink hints---arithmetic operations for CWE-190 and CWE-191, free-like operations for CWE-415 and CWE-416---and asks the model to follow a fixed reasoning checklist: identify the dangerous operation, trace the origin of dangerous data or pointer state, check for validation evidence before the sink, analyze whether control-flow evidence shows that validation guards the sink, and classify the function as \texttt{VULNERABLE} or \texttt{SAFE}. The checklist is identical across all representation variants so that the representation is the sole experimental variable.

\subsubsection{Dynamic Few-Shot Selection}

The prompt generator supports dynamic few-shot selection. For each CWE, the generator identifies available examples matched to the target CWE, estimates the space required by the target representation, and includes only those examples that fit within the remaining prompt budget. When no matched examples fit, the prompt explicitly notes that no examples are included. This policy prioritizes the representation evidence for the evaluated testcase over auxiliary few-shot context---a choice that is itself an experimental confound, discussed further in \Cref{sec:threats}.

\subsubsection{Prompt Budget Control}

The benchmark enforces a fixed maximum prompt size of 50,000 characters in all reported experiments. We chose a fixed character budget to enforce a consistent prompt-size constraint across variants and to expose representation-level differences in prompt pressure under a uniform ceiling. In addition to character counts, we report input-token counts returned by the tokenizer of the evaluated \texttt{qwen3.6-plus} service. Prompts were reconstructed from the retained testcase and representation artifacts and audited against the historical \texttt{prompt\_chars} values: 918 of 1,068 prompts matched exactly, while 150 older prompts used the closest reproducible template and are marked \texttt{proxy\_reconstructed} in the artifact. Exact-only sensitivity checks preserve the family ordering and change family averages by at most 3.0\%. For every prompt, the system records original prompt size, final prompt size, and whether clipping occurred. When the initial prompt exceeds the budget, the generator first removes few-shot examples; if the prompt remains over budget, the remaining character budget is distributed across content blocks proportionally and each over-budget block is truncated with an explicit marker. Clipping is treated as an explicit experimental factor rather than an invisible preprocessing step.

\subsubsection{Structured Output}

The LLM is required to return a structured JSON object with a reasoning chain, binary conclusion, confidence score, CWE mapping, and brief explanation:

\begin{lstlisting}
{
  "reasoning": {
    "sink_identified":      "line and operation",
    "data_source":          "origin of dangerous data or pointer state",
    "validation_present":   true,
    "control_flow_analysis":"whether validation guards the sink",
    "pdg_dependencies":     ["relevant DDG/CDG evidence"]
  },
  "conclusion":   "VULNERABLE or SAFE",
  "confidence":   0.0,
  "cwe_mapping":  "CWE identifier",
  "explanation":  "brief explanation"
}
\end{lstlisting}

The \texttt{conclusion} field is normalized to \texttt{VULNERABLE} or \texttt{SAFE}. Structured output enables automatic accuracy computation, representation-level comparison, and downstream error analysis.

\section{Experimental Setup}\label{sec:setup}

\subsection{Evaluation Conditions}

The main experiment is the Joern-based representation ablation. For each of the 107 standard testcases, the benchmark evaluates  
up to 
ten prompt variants listed in \Cref{tab:variants}, totaling 1,068 attempted evaluation rows.
All Joern-based experiments use \texttt{full\_file\_target\_method} as the scope setting.

The ePDG track is treated as a complementary external-graph evaluation. Each retained ePDG testcase contributes one ePDG prompt. Because ePDG generation has separate feasibility constraints, ePDG results are not pooled with the standard Joern ablation except when direct paired comparison is performed on the 11-case overlapping subset.

\subsection{Model and Inference Configuration}

All experiments use \texttt{qwen3.6-plus}~\cite{qwen36plus} as the inference model, accessed through a JSON-formatted chat-completion interface.\footnote{The exact API model identifier used in this study was \texttt{qwen3.6-plus}. Researchers replicating these experiments should verify the current model identifier with the API provider, as naming conventions may vary across API versions.} The sampling temperature is fixed at $T{=}0.2$ and the maximum response length is 2048 tokens. These settings are held constant across all variants to prevent model-side confounding. Each testcase-variant pair receives exactly one inference query; replication with multiple runs is left for future work and noted as a threat in \Cref{sec:threats}.

\subsection{Evaluation Procedure}

For each testcase and each applicable representation variant, the benchmark constructs a single prompt and executes one LLM-based vulnerability classification query. The expected label is binary: vulnerable functions map to \texttt{VULNERABLE}, fixed functions map to \texttt{SAFE}. A prediction is counted as correct if the normalized model conclusion exactly matches the expected label.

The system records the final prediction, reasoning fields, prompt statistics, clipping status, and timing metadata. Failed requests, unparsable outputs, duplicate testcase-variant records, and intermediate reruns are removed from the final retained result sets. Model confidence and inference latency are logged in the result files for future analysis.

\subsection{Metrics}

We report two complementary accuracy metrics.

\paragraph{Effective accuracy.}
This metric treats failed, unknown, or unparsable responses as end-to-end classification failures. It reflects practical deployment behavior, where any non-answer is an error.

\paragraph{Curated accuracy.}
Curated accuracy evaluates correctness only on retained rows that produced valid binary predictions. This metric isolates representation quality from response-format failures.

Both metrics are reported because they capture different phenomena: effective accuracy measures end-to-end pipeline reliability, while curated accuracy measures the quality of representations for cases where the model does respond. 
Since our study focuses primarily on examining the impact of program representations, \textit{we mean the curated accuracy when referring to ``accuracy" without qualification}.
We also report prompt-overhead indicators (average prompt size in characters and input tokens, together with clipping frequency) as practical measures of representation cost.

\section{Results} \label{sec:results}
In this section, we present and discuss our experimental results for each RQ, followed by 
in-depth case studies that provide explanatory evidence supporting some of the key RQ2–RQ4 findings.

\subsection{Overall Effectiveness of Graphs vs. raw source (RQ1)}
The main Joern evaluation track contains 107 standard testcases evaluated under up to ten prompt variants, yielding 1,068 attempted evaluation rows. Among these, 1,015 produced valid binary predictions, while 53 resulted in \texttt{UNKNOWN} outputs. Across all standard-track rows, the benchmark achieved 67.3\% effective accuracy and 70.8\% curated accuracy. The gap between effective and curated accuracy indicates that most model outputs were parseable and usable, but response reliability remains a non-negligible part of end-to-end LLM-based vulnerability reasoning.

\begin{table}[t]
\centering
\caption{Aggregate effectiveness by prompt family. Graph-only prompts achieve the highest accuracy.}
\label{tab:family}
\setlength{\tabcolsep}{7pt}
\begin{tabular}{lrrrrr}
\toprule
\rowcolor{headergray}
\hdr{Family} & \hdr{Attempted} & \hdr{Retained} & \hdr{Correct} & \hdr{Eff.\ Acc.} & \hdr{Cur.\ Acc.} \\
\midrule
Source-only   & 107   & 101   & 54  & 50.5\%  & 53.5\% \\
\rowcolor{bestgreen}
Graph-only    & 642   & 613   & 456 & 71.0\%  & \textbf{74.4\%} \\
Source+Graph  & 319   & 301   & 209 & 65.5\%  & 69.4\% \\
\midrule
Total         & 1,068 & 1,015 & 719 & 67.3\%  & 70.8\% \\
\bottomrule
\end{tabular}
\end{table}

\Cref{tab:family} shows a family-level contrast. The raw source-code baseline achieves 53.5\% curated accuracy, while the graph-only family reaches 74.4\%, a 20.9 percentage-point (pp) improvement. The same pattern holds under effective accuracy, where graph-only prompts improve over source-only prompts by 20.5 pp. Thus, the advantage of structural representations is not an artifact of excluding \texttt{UNKNOWN} outputs; it appears in both the end-to-end and retained-output metrics.

This result provides the first answer to RQ1: standalone structural graph representations can support LLM-based vulnerability reasoning substantially better than raw source-code prompting in this corpus. The comparison is especially notable because graph-only prompts do not include the original source text as a separate input. Instead, they present the model with serialized structural evidence extracted from static analysis. This suggests that explicit program structure can substitute for, and in this setting outperform, direct source-code prompting.

The source-plus-graph family occupies an intermediate position, with 69.4\% curated accuracy. Although these prompts contain strictly more information than graph-only prompts, they do not achieve higher accuracy. This result previews the context-dilution pattern analyzed later: adding raw source context to structural evidence may introduce irrelevant or misleading information that weakens the model's focus on vulnerability-relevant relations. At this point, however, the family-level result should be interpreted conservatively: \Cref{tab:family} pools multiple variants within each family, so the next subsection examines which individual representations and combinations are responsible for the observed gains.

\begin{rqfinding}{Finding 1 (RQ1)}
In the standard Joern track, standalone graph-based prompts substantially outperform raw source-code prompting. The graph-only family achieves 74.4\% curated accuracy compared with 53.5\% for source-only prompts, indicating that compact structural program evidence can be more effective than raw source text for LLM vulnerability reasoning.
\end{rqfinding}

\vspace{-4pt}
\subsection{Effectiveness Across Representation Variants (RQ2)}

\begin{table}[t]
\centering
\caption{Detailed effectiveness by representation variant, sorted by curated accuracy. The most accurate variant is highlighted.}
\label{tab:variant-results}
\setlength{\tabcolsep}{7pt}
\begin{tabular}{lrrrrr}
\toprule
\rowcolor{headergray}
\hdr{Variant} & \hdr{Attempted} & \hdr{Retained} & \hdr{Correct} & \hdr{Eff.\ Acc.} & \hdr{Cur.\ Acc.} \\
\midrule
\rowcolor{bestgreen}
\texttt{ast\_pdg}          & 107 & 101 & 84 & 78.5\% & \textbf{83.2\%} \\
\texttt{ast\_cfg}          & 107 & 102 & 82 & 76.6\% & 80.4\% \\
\texttt{cfg\_pdg}          & 107 & 103 & 76 & 71.0\% & 73.8\% \\
\texttt{full}              & 106 & 98  & 72 & 67.9\% & 73.5\% \\
\texttt{ast}               & 107 & 104 & 75 & 70.1\% & 72.1\% \\
\texttt{ast\_plus\_source} & 106 & 101 & 70 & 66.0\% & 69.3\% \\
\texttt{pdg}               & 107 & 100 & 69 & 64.5\% & 69.0\% \\
\texttt{cfg}               & 107 & 103 & 70 & 65.4\% & 68.0\% \\
\texttt{pdg\_plus\_source} & 107 & 102 & 67 & 62.6\% & 65.7\% \\
\texttt{raw}               & 107 & 101 & 54 & 50.5\% & 53.5\% \\
\bottomrule
\end{tabular}
\end{table}

\Cref{tab:variant-results} reveals a clear accuracy hierarchy. The most accurate variant is \texttt{ast\_pdg}, achieving 83.2\% curated accuracy---a 29.7 percentage point improvement over the raw-source baseline. \texttt{ast\_cfg} follows at 80.4\%, and \texttt{cfg\_pdg} at 73.8\%.

\vspace{-4pt}
\paragraph{Why AST+PDG outperforms other combinations.}
The superiority of \texttt{ast\_pdg} over \texttt{ast\_cfg} and \texttt{cfg\_pdg} can be understood through the complementary roles of the constituent representations. ASTs provide local syntactic anchors---expression structure, operator types, call sites, and variable declarations---that ground the model's understanding of what operations occur at each line. PDGs expose cross-statement dependency relations, making explicit which values flow into which sinks and which control conditions guard which operations. Together, they provide both local syntactic detail and semantic dependency evidence, the combination most directly relevant to the source-to-sink reasoning required for the selected vulnerability classes.

In contrast, \texttt{ast\_cfg} combines syntactic structure with execution-order information but lacks explicit data-dependence edges; it performs well (80.4\%) because guard structures and control paths are important for the arithmetic and memory-lifecycle bugs in the corpus, but misses some dependency-propagation evidence that PDGs capture. \texttt{cfg\_pdg} (73.8\%) combines path and dependency evidence but lacks AST's local syntactic detail, which may make it harder to anchor reasoning to specific expressions.

\vspace{-4pt}
\paragraph{Source-plus-graph variants underperform.}
Notably, adding raw source code to graph evidence does not help. \texttt{pdg\_plus\_source} (65.7\%) underperforms the standalone \texttt{pdg} (69.0\%), and \texttt{ast\_plus\_source} (69.3\%) underperforms \texttt{ast\_pdg} (83.2\%). The maximum-information condition \texttt{full} achieves only 73.5\%, below both two-graph-only variants. This pattern is consistent with context dilution, explored further in \Cref{sec:overhead} and \Cref{sec:casestudies}.

\begin{rqfinding}{Finding 2 (RQ2)}
Multi-view graph-only combinations outperform single-graph variants. AST+PDG achieves the highest curated accuracy (83.2\%) by combining local syntactic anchors from the AST with cross-statement dependency evidence from the PDG. AST+CFG (80.4\%) is the next most accurate, while CFG+PDG (73.8\%) trails due to the absence of AST's local syntactic grounding.
\end{rqfinding}

\subsection{Representation Cost and Cost-Effectiveness Tradeoff (RQ3)}
\label{sec:overhead}

\begin{table}[t]
\centering
\caption{Prompt cost and clipping by representation family. Input-token counts use the evaluated model's tokenizer.}
\label{tab:overhead}
\setlength{\tabcolsep}{9pt}
\begin{tabular}{lrrr}
\toprule
\rowcolor{headergray}
\hdr{Family} & \hdr{Avg.\ Chars} & \hdr{Avg.\ Input Tokens} & \hdr{Clipped Prompts} \\
\midrule
Source-only   & 34,666 & 10,767 & 47 (43.9\%) \\
\rowcolor{bestgreen}
Graph-only    & 16,858 & 5,973  & 69 (10.7\%) \\
Source+Graph  & 40,804 & 12,938 & 191 (59.9\%) \\
\bottomrule
\end{tabular}
\end{table}

\Cref{tab:overhead} first isolates prompt overhead from classification effectiveness. Graph-only prompts are the most compact family, averaging 16,858 characters and 5,973 input tokens per prompt. Source-only prompts are roughly twice as large, averaging 34,666 characters and 10,767 tokens. Source-plus-graph prompts are the largest, averaging 40,804 characters and 12,938 tokens. Thus, source-plus-graph prompts require 2.4$\times$ as many characters and 2.2$\times$ as many input tokens as graph-only prompts.

Because the three families contain different numbers of attempted prompts, clipping rates are more informative than raw clipping counts. Graph-only prompts have the lowest clipping rate, at 10.7\% (69 of 642 prompts), compared with 43.9\% for source-only prompts and 59.9\% for source-plus-graph prompts. Source-plus-graph therefore imposes the greatest prompt-budget pressure: it is both the largest family by average input size and the most frequently clipped family by rate.

Graph-only still contains 69 clipped prompts despite having the lowest family-level clipping rate. This reflects representation heterogeneity: PDG outputs can grow large for functions with many dependencies, while AST and CFG outputs tend to be smaller. 

This suggests that graph-only prompting is already more compact at the family level, but could be further improved through 
more fine-grained prompt-budget management (e.g., variant-aware graph pruning, dependence slicing, or edge summarization).

\begin{rqfinding}{Finding 3 (RQ3)}
Graph-only representations impose substantially lower prompt overhead than both raw source and source-plus-graph prompts. They use the fewest characters and input tokens on average and have the lowest clipping rate, indicating that structural graph representations can reduce prompt-budget pressure rather than merely adding analysis metadata to the prompt.
\end{rqfinding}

\paragraph{Tradeoff between effectiveness and cost.}

\begin{figure}[t]
    \centering
    \includegraphics[width=\linewidth]{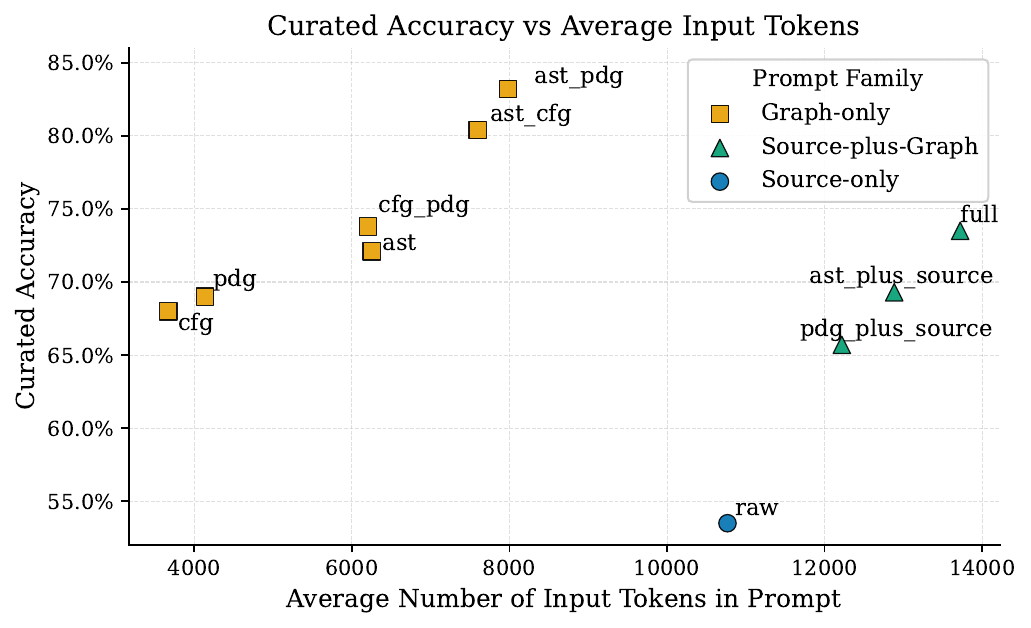}
    \caption{Curated accuracy versus average input-token count for each representation variant. Colors and markers indicate prompt families.}
    \label{fig:accvspromptsize}
\end{figure}

\Cref{fig:accvspromptsize} shows no monotonic relationship between input-token count and curated accuracy. The raw-source baseline is relatively costly, averaging 10,767 input tokens, yet achieves the lowest curated accuracy at 53.5\%. In contrast, \texttt{ast\_pdg} uses only 7,980 input tokens on average while achieving the highest accuracy at 83.2\%. The three source-plus-graph variants are the largest conditions, averaging approximately 12,200--13,700 input tokens, but achieve only 65.7\%--73.5\% curated accuracy. These results indicate that representation content and focus matter more than input volume alone.

The family-level comparison reinforces this pattern. Graph-only prompts combine the lowest average input cost (5,973 tokens) with the highest curated accuracy (74.4\%). Source-plus-graph prompts average 12,938 tokens but reach only 69.4\% curated accuracy, while source-only prompts average 10,767 tokens and reach 53.5\%. More input therefore does not imply greater classification effectiveness; the type and concentration of evidence determine whether additional context is useful.

\begin{rqfinding}{Finding 4 (RQ3)}
Graph-only representations provide the best effectiveness--cost tradeoff: they achieve 74.4\% curated accuracy at an average of 16,858 characters (5,973 input tokens), while source-plus-graph prompts require 2.4$\times$ more characters and 2.2$\times$ more input tokens yet achieve 5.0 percentage points lower curated accuracy (69.4\% versus 74.4\%). Representation choice thus offers a lever for simultaneously improving accuracy and reducing prompt overhead.
\end{rqfinding}

\begin{ctxfinding}{Finding 5 (RQ3)---Context Dilution Effect}
These results suggest that larger prompts do not inherently produce better vulnerability reasoning. Source-plus-graph prompts are larger than graph-only prompts but perform worse, consistent with the hypothesis that additional raw source context dilutes the vulnerability-relevant signal. This effect appears distinct from simple truncation: as shown in Case Study~2 (\Cref{sec:casestudies}), source-augmented prompts fail even when no clipping occurs. The mechanism appears to be semantic distraction rather than information loss.
\end{ctxfinding}

\subsection{Variation Across CWE Categories, Labels, and ePDG Inputs (RQ4)}
We further break down the aggregate performance across vulnerability types and look into several important factors that also 
impact our interpretation of representation effects. 

\subsubsection{Effectiveness by CWE Category}
\Cref{tab:agg_cwe} reports curated accuracy for the raw baseline, the strongest individual variant, and the two graph-containing families. CWE-190 contributes 56 of the 107 testcases and therefore has the greatest influence on the aggregate results. CWE-122 and CWE-191 contain only four and two testcases, respectively, and are marked low-support; their percentages are descriptive and should not be used for strong category-level conclusions.

The benefit of structural representations is not limited to the CWE-190 majority. Among the adequately supported categories, AST+PDG improves over raw source by 30.8 percentage points on CWE-190, 22.7 points on CWE-415, and 33.3 points on CWE-416. The pooled graph-only family likewise exceeds raw source by 24.0, 14.6, and 20.9 points, respectively. These results indicate that the aggregate graph advantage appears across both arithmetic and memory-lifecycle vulnerabilities rather than being solely an artifact of the CWE-190-heavy distribution.

\begin{rqfinding}{Finding 6 (RQ4)}
The advantage of structural representations is not confined to the dominant CWE-190 category. On the three adequately supported CWE categories, AST+PDG improves over raw source by 30.8 percentage points on CWE-190, 22.7 points on CWE-415, and 33.3 points on CWE-416. This suggests that the aggregate graph advantage reflects a broader representation effect across both arithmetic and memory-lifecycle vulnerabilities, not merely the skew toward CWE-190.
\end{rqfinding}

The relative ordering nevertheless varies by CWE. Source-plus-graph performs best on CWE-415 at 76.2\%, exceeding AST+PDG at 72.7\% and the graph-only family at 64.6\%, while it underperforms graph-only substantially on CWE-190 (65.8\% versus 77.8\%). One possible explanation is that raw source context can help distinguish repeated cleanup and pointer-lifecycle operations, whereas focused dependency evidence is more useful for arithmetic source-to-sink reasoning. Model familiarity with particular vulnerability patterns and the CWE-specific guidance in each prompt may also contribute to the variation. These interpretations remain provisional, and larger, better-balanced per-CWE experiments are needed before drawing category-specific conclusions.

\begin{table}[t]
\centering
\caption{Curated accuracy by CWE for major variants and families. Family columns pool retained predictions across variants in that family.}
\label{tab:agg_cwe}
\setlength{\tabcolsep}{8pt}
\begin{tabular}{lrrrrr}
\toprule
\rowcolor{headergray}
\hdr{CWE} & \hdr{Cases} & \hdr{Raw} & \hdr{AST+PDG} & \hdr{Graph-only} & \hdr{Source+Graph} \\
\midrule
CWE-122$^\dagger$ & 4  & 50.0\%  & 100.0\% & 83.3\% & 58.3\% \\
CWE-190             & 56 & 53.8\%  & 84.6\%  & 77.8\% & 65.8\% \\
CWE-191$^\dagger$ & 2  & 100.0\% & 100.0\% & 83.3\% & 50.0\% \\
CWE-415             & 23 & 50.0\%  & 72.7\%  & 64.6\% & 76.2\% \\
CWE-416             & 22 & 52.4\%  & 85.7\%  & 73.3\% & 75.4\% \\
\bottomrule
\end{tabular}
\vspace{2pt}

\raggedright\footnotesize{$^\dagger$Low support ($<5$ testcases); percentages are descriptive only. Curated accuracy excludes \texttt{UNKNOWN} predictions.}
\end{table}

\begin{figure}[t]
    \centering
    \includegraphics[width=\linewidth]{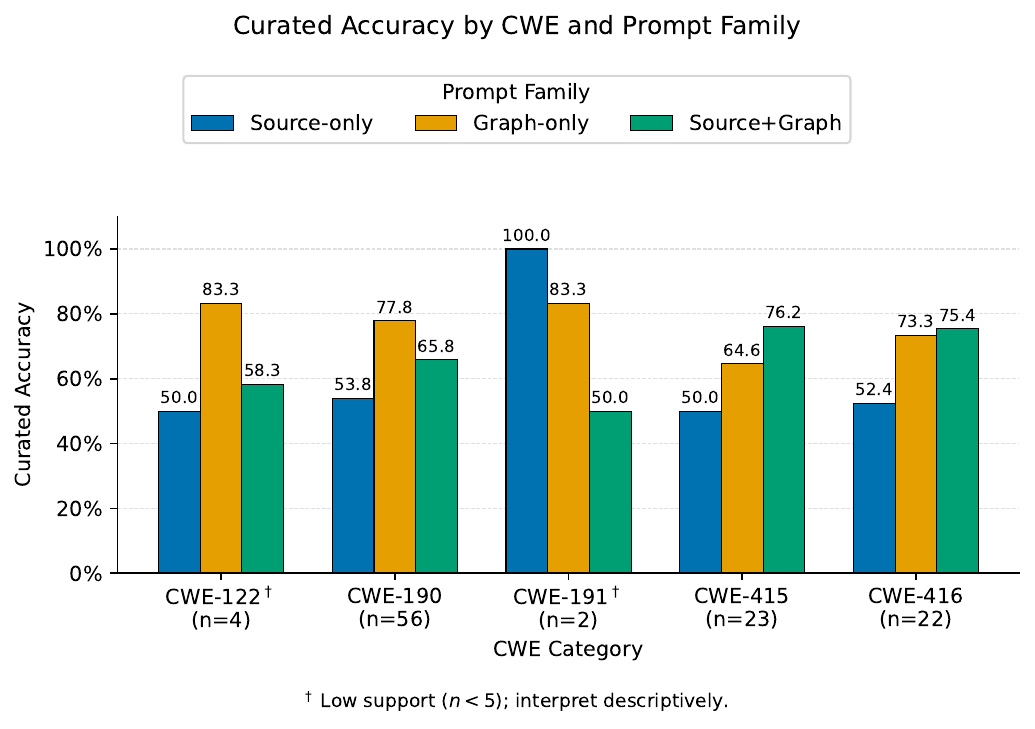}
    \caption{Curated accuracy versus CWE category grouped by prompt family. Colors indicate prompt families. Poorly supported CWE categories are denoted.}
    \label{fig:accvscwe_promptfam}
\end{figure}

\begin{figure}[t]
    \centering
    \includegraphics[width=\linewidth]{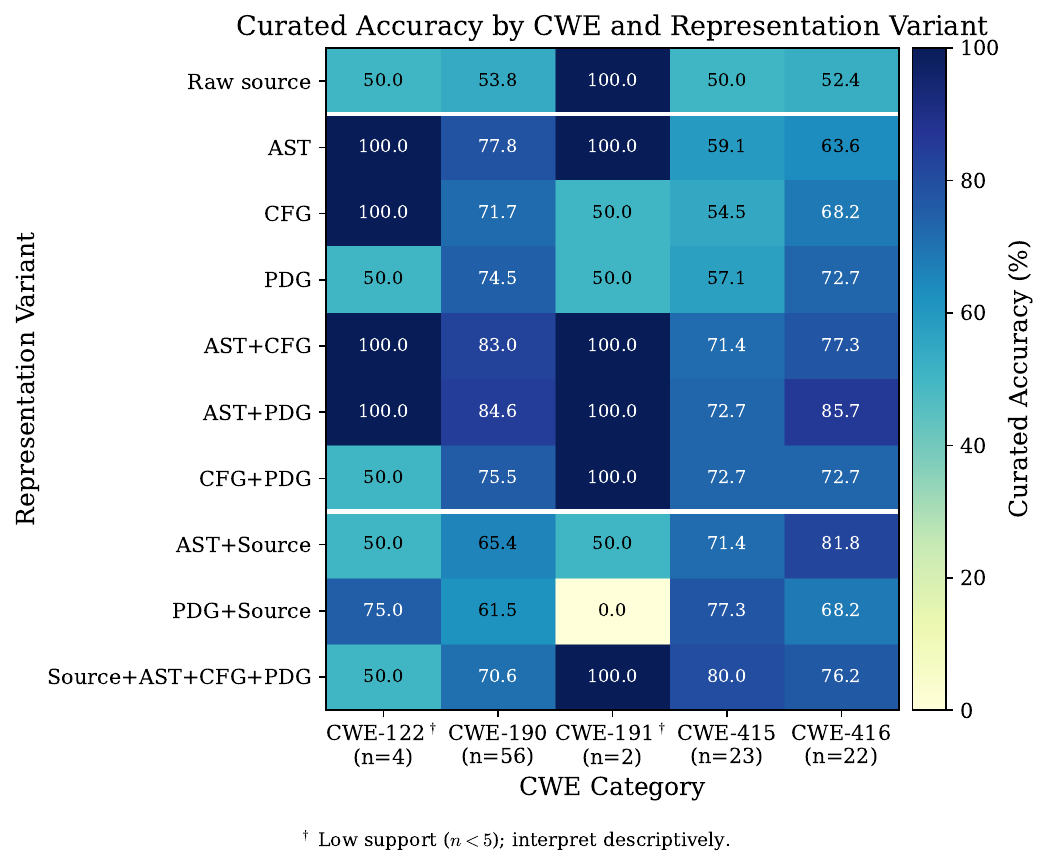}
    \caption{Curated accuracy by CWE category and representation variant. Cell color indicates curated accuracy. Low-support CWE categories are marked and should be interpreted descriptively.}
    \label{fig:accvscwe_variant}
\end{figure}

In \Cref{fig:accvscwe_promptfam} and \Cref{fig:accvscwe_variant}, the two CWE breakdown figures provide complementary views of the same trend. 
At the variant level, structured graph representations outperform the raw-source baseline on the three adequately supported categories: CWE-190 ($n{=}56$), CWE-415 ($n{=}23$), and CWE-416 ($n{=}22$). 
Raw source remains near 50--54\% curated accuracy, while the strongest graph variants reach 84.6\% on CWE-190, 72.7\% on CWE-415, and 85.7\% on CWE-416. 
The heatmap suggests that combined graph views are often strongest, especially AST+PDG for CWE-190 and CWE-416, although the best representation varies by category.

The prompt-family view shows that source context contributes differently across CWEs. 
For CWE-190, graph-only prompts outperform source-plus-graph prompts (77.8\% versus 65.8\%), consistent with the broader context-dilution pattern. 
For memory-lifecycle categories, however, source-plus-graph is best on CWE-415 (76.2\%) and narrowly leads graph-only on CWE-416 (75.4\% versus 73.3\%), suggesting that token-level source context may help with cleanup order and pointer-lifetime evidence. 
CWE-122 and CWE-191 remain descriptive only because of their low support ($n{=}4$ and $n{=2}$).

\begin{rqfinding}{Finding 7 (RQ4)}
Representation effects vary by vulnerability category. Graph-only prompts are strongest for CWE-190, consistent with the value of focused structural evidence for arithmetic source-to-sink reasoning. In contrast, source-plus-graph prompts perform competitively or best on memory-lifecycle categories such as CWE-415 and CWE-416, suggesting that raw source context may still help when reasoning about cleanup order, repeated frees, and pointer-lifetime evidence. Thus, graph representations are broadly useful, but the best representation strategy may depend on the vulnerability class.
\end{rqfinding}

\subsubsection{Label-Wise Behavior}

The benchmark exhibits a label-wise asymmetry across the full standard corpus (520 SAFE rows, 548 VULNERABLE rows). Effective accuracy on SAFE rows reaches 71.0\%, with curated accuracy of 76.1\%. In contrast, effective accuracy on VULNERABLE rows is 63.9\%, with curated accuracy of 66.0\%. The model is therefore more reliable at recognizing explicit safety evidence than at identifying all vulnerability evidence.

This asymmetry may reflect multiple factors: SAFE functions may contain more explicit validation patterns that the model can identify as protective, whereas VULNERABLE functions may require the model to reason about a subtle absence of a guard or a non-obvious data-flow path. The asymmetry could also reflect conservative model behavior under uncertainty, wherein the model preferentially outputs \texttt{SAFE} when evidence is ambiguous. Future work should analyze whether this pattern is driven by dataset characteristics, representation information loss, model conservatism, or prompt design.

\begin{rqfinding}{Finding 8 (RQ4)}
The model performs better on \texttt{SAFE} cases than on \texttt{VULNERABLE} cases across the standard corpus: curated accuracy is 76.1\% for \texttt{SAFE} rows versus 66.0\% for \texttt{VULNERABLE} rows. This suggests that the current prompting and representation design more reliably recognizes explicit safety evidence than subtle vulnerability evidence, such as missing guards, unsafe value propagation, or non-obvious pointer-lifecycle errors.
\end{rqfinding}

\subsubsection{Auxiliary ePDG Evaluation}

\begin{table}[t]
\centering
\caption{Auxiliary ePDG evaluation track statistics. Results are not directly comparable to the 107-case Joern track except on the 11-case overlapping subset.}
\label{tab:epdg}
\setlength{\tabcolsep}{10pt}
\begin{tabular}{lr}
\toprule
\rowcolor{headergray}
\hdr{Statistic} & \hdr{Value} \\
\midrule
Retained ePDG cases              & 19     \\
Direct paired-comparison subset  & 11     \\
Correct ePDG predictions         & 11     \\
Curated accuracy                 & 57.9\% \\
Average prompt chars             & 40,289 \\
Prompt clipping count            & 2      \\
\texttt{function\_only} cases    & 13     \\
\texttt{full\_file} cases        & 6      \\
\bottomrule
\end{tabular}
\end{table}

The auxiliary ePDG track achieves 57.9\% curated accuracy over 19 retained cases. Prompt clipping occurs only twice, indicating that retained ePDG inputs can remain manageable despite dense dependence information. The 57.9\% figure should not be directly compared with the standard Joern-track results: the testcases are different, the corpus is smaller, and the representation format differs substantially. On the 11-case overlapping subset, direct comparison is feasible but limited in statistical power.

The lower accuracy relative to most Joern variants may reflect several factors. ePDG records encode dependence information at the LLVM IR level, where variable names, types, and code labels are more abstract and less immediately readable than C/C++ source-level constructs. The format may therefore present an additional comprehension challenge for the model. At the same time, the low clipping rate (2 of 19 cases) suggests that prompt size is not the primary limiting factor in the ePDG track---the issue is more likely representation readability and formatting rather than context length.

\begin{epdgfinding}{Finding 9 (RQ4)---ePDG effects}
The auxiliary ePDG track provides preliminary evidence that vulnerability-oriented dependence metadata can support LLM reasoning, but current ePDG prompts remain limited by representation readability and artifact-generation feasibility rather than prompt length. With only 19 retained cases and 57.9\% curated accuracy, these results should be treated as auxiliary evidence rather than a direct comparison with the standard Joern-track results.
\end{epdgfinding}

\subsection{Case Studies}
\label{sec:casestudies}

\paragraph{Case Study 1 (Graph-only success, Source-only failure).}

\begin{table}[t]
  \centering
  \caption{Attributes of testcase for reproducibility.}
  \label{tab:casestudy1_reproduciblity}
  \setlength{\tabcolsep}{12pt}
  \begin{tabularx}{\linewidth}{X}
    \toprule
    \rowcolor{headergray}
    \hdr{Case Study 1 \& 2 Testcase Attributes}\\
    \midrule
    \textbf{Testcase ID:} FreeRDP\_\_CWE-190\_\_msusb\_c\_\_9f77fc3dd239\_\_cse713\_fixed \\
    \textbf{Project name:} FreeRDP \\
    \textbf{Project file:} msusb.c\\
    \textbf{Patch commit:} 9f77fc3dd2394373e1be753952b00dafa1a9b7da \\
    \textbf{CWE:} CWE-190 \\ 
    \textbf{Expected label:} SAFE \\
    \textbf{Representation outcomes:} \texttt{raw}: VULNERABLE (incorrect); \texttt{ast\_pdg}: SAFE (correct) \\
    \textbf{Case 2 outcomes:} all graph-only: SAFE (correct); all source-plus-graph: VULNERABLE (incorrect) \\
    \bottomrule
  \end{tabularx}
\end{table}

The fixed FreeRDP~\cite{FreeRDP} integer-overflow testcase in function \texttt{msusb\_mspipes\_read} illustrates how compact graph evidence can distinguish guarded from unguarded arithmetic. The source-only baseline (\texttt{raw}) predicted \texttt{VULNERABLE}, whereas \texttt{ast\_pdg} correctly predicted \texttt{SAFE} with confidence 0.95. The relevant source fragment reads:

\begin{lstlisting}
if (Stream_GetRemainingCapacity(s) / 12 < NumberOfPipes)
    return NULL;
MsPipes = calloc(NumberOfPipes, sizeof(...));
\end{lstlisting}

The AST identified the comparison operator, the early-return statement, the \texttt{calloc} call, and the loop structure surrounding the allocation. The PDG linked \texttt{NumberOfPipes} and the stream-capacity expression to the guard at line~67 and then to the later allocation at line~70. The model's reasoning for \texttt{ast\_pdg} explicitly identified the guard as bounding \texttt{NumberOfPipes} by available stream capacity before the allocation sink executes. The model's reasoning for \texttt{raw} fixated on the integer division rather than recognizing it as a guard, misclassifying the function as vulnerable. This case illustrates why compact graph evidence can help distinguish guarded arithmetic from unguarded arithmetic---a distinction that is easy to miss when the guard and sink appear in a larger block of source text.

\paragraph{Case Study 2 (Source-augmented failure without truncation).}

The same FreeRDP testcase illustrates the context dilution effect independently of prompt truncation. All six graph-only conditions predicted \texttt{SAFE} correctly, but all three source-augmented conditions (\texttt{full}, \texttt{ast\_plus\_source}, \texttt{pdg\_plus\_source}) predicted \texttt{VULNERABLE} incorrectly. Crucially, this failure was \emph{not} caused by clipping: the source-augmented prompts ranged from 17,853 to 26,008 characters and remained well within the 50,000-character budget. The failure arose because the source-augmented prompts exposed additional file-level source context surrounding the target method, including a separate interface-writing routine that computes a buffer size as \texttt{16 + NumberOfPipes * 20}. The model attended to this unrelated computation and treated it as the vulnerability sink. The graph-only prompts were more tightly scoped to the target-method evidence, keeping the line-67 guard and line-70 allocation prominent. This case demonstrates that context dilution can occur at the semantic level---extra source text causes the model to focus on the wrong code---even when prompt length is well below the budget.

\paragraph{Case Study 3 (Auxiliary ePDG behavior).}

\begin{table}[t]
  \centering
  \caption{Attributes of testcase for reproducibility.}
  \label{tab:casestudy3_reproduciblity}
  \setlength{\tabcolsep}{12pt}
  \begin{tabularx}{\linewidth}{X}
    \toprule
    \rowcolor{headergray}
    \hdr{Case Study 3 Testcase Attributes}\\
    \midrule
    \textbf{Testcase ID:} openexr\_\_CWE-190\_\_ImfFastHuf\_cpp\_\_c3ed4a1db1f3\_\_cse713\_vulnerable \\
    \textbf{Project name:} openexr \\
    \textbf{Project file:} ImfFastHuf.cpp\\
    \textbf{Patch commit:} c3ed4a1db1f39bf4524a644cb2af81dc8cfab33f \\
    \textbf{CWE:} CWE-190 \\ 
    \textbf{Representation:} ePDG \\
    \textbf{Expected label:} VULNERABLE \\
    \textbf{Representation outcome:} VULNERABLE (correct; confidence 0.85) \\
    \bottomrule
  \end{tabularx}
\end{table}

In a vulnerable OpenEXR~\cite{OpenEXR} integer-overflow testcase from \texttt{ImfFastHuf}, the ePDG prompt correctly predicted \texttt{VULNERABLE} with confidence 0.85. The retained ePDG evidence annotated arithmetic operations at lines~130 and~147 as candidate root-cause nodes and exposed post-arithmetic comparison edges. Specifically, the serialized ePDG linked subtraction operations to later integer-comparison nodes through data-dependence edges and tagged the line-130 subtraction as a root-cause arithmetic operation. The model's reasoning distinguished these post-arithmetic checks from true precondition guards: comparisons that occur \emph{after} the arithmetic do not prevent an overflow that has already occurred. This case illustrates the potential value of vulnerability-oriented dependence metadata. At the same time, the auxiliary ePDG track also produced false positives, such as a fixed \texttt{libgd} CWE-415 testcase in which the model over-interpreted repeated cleanup calls as double-free evidence. This reinforces why ePDG results are reported as auxiliary evidence pending future improvements in artifact generation reliability and representation readability.

\vspace{4pt}\noindent
\textbf{Case-Study Takeaway.}
The case studies explain the mechanisms behind the aggregate results. Case Study~1 shows that AST+PDG can make guard and dependence evidence more explicit than raw source. Case Study~2 shows that source augmentation can fail through semantic distraction even without prompt clipping. Case Study~3 shows that ePDG metadata can help identify vulnerability-specific dependence patterns, but its usefulness remains limited by representation readability and occasional over-interpretation.

\section{Discussion}

\subsection{The Context Dilution Effect}

The most important finding of this study is that more input information does not necessarily improve LLM vulnerability reasoning. Source-plus-graph prompts are substantially larger than graph-only prompts and include both raw code and structural evidence, yet they consistently underperform at the family level. Our results suggest two distinct mechanisms through which context dilution can arise.

The first mechanism is \emph{prompt-length pressure}: when prompts approach or exceed the context budget, few-shot examples are removed and content may be clipped, reducing the quality of in-context guidance. This is evidenced by the high clipping frequency (191 cases) in the source-plus-graph family.

The second mechanism is \emph{semantic distraction}: even when the full augmented prompt fits within the budget, additional source context can cause the model to attend to wrong code, as demonstrated in Case Study~2. The model in that case focused on a buffer-size computation in an unrelated interface routine rather than on the guard-protected allocation in the target method. This failure occurred without any truncation, indicating that the distraction arose from raw source content competing for model attention with the graph-derived structural evidence.

The implication is that graph-only representations are not merely smaller than source-augmented prompts---they are also more focused, preserving only the structural evidence most relevant to the vulnerability question.

\subsection{Why Graph Representations Help}

Graph representations appear to improve LLM vulnerability reasoning through three related mechanisms. First, they expose structural relations that are implicit in source code: a PDG can directly link a value definition to a later dangerous use, while raw source requires the model to reconstruct that relation from syntax alone. Second, they reduce syntactic noise: formatting artifacts, comments, helper declarations, and irrelevant statements are suppressed or compressed, leaving the structural evidence more prominent. Third, they provide semantic indexing through line numbers, node types, edge labels, and dependency summaries that give the model explicit anchors for reasoning about the target vulnerability.

The superior performance of AST+PDG is consistent with this analysis. The AST provides statement-level and expression-level structure---the \emph{what} of each operation---while the PDG provides the \emph{how values reach sinks}. Together they offer a compact but sufficiently expressive view of the function for the vulnerability classes studied.

It is important to note that graph representations are not universally better than source code by construction. Their value depends on the quality and completeness of the graph extraction, the fidelity of the serialization, and the extent to which the target vulnerability class exposes evidence through structural rather than lexical cues. Future work should examine vulnerability classes where raw source may retain advantages.

\subsection{Representation Efficiency and Static Analysis as a Prompt-Construction Layer}

Representation efficiency matters in deployment: longer prompts incur higher latency, greater inference cost, and increased risk of truncation. In this study, graph-only prompts achieve the best accuracy while requiring the fewest characters on average. This means that representation engineering can simultaneously improve both effectiveness and efficiency.

These results suggest that future LLM-based software analysis systems should treat static analysis not only as a standalone detection method but also as a \emph{prompt-construction layer}: a mechanism for extracting compact, security-relevant structural evidence that LLMs can reason about more reliably than over unfiltered source code. This framing connects program analysis research to prompt engineering and suggests a productive design space for representation-aware LLM security tools.

\subsection{Implications for LLM-Based Software Analysis}

Beyond vulnerability detection, the context dilution effect and the value of compact structural representations have implications for other LLM-based software analysis tasks, including patch analysis, secure code review, and agentic code inspection. Tasks where security-relevant evidence is sparse relative to total code context---as is typical in real-world projects---may particularly benefit from structured representation. This paper provides an initial empirical foundation for such representation-aware reasoning.

\section{Threats to Validity}
\label{sec:threats}

\subsection{Single Model}

The current study uses a single inference model, \texttt{qwen3.6-plus}. Accuracy values may not generalize to other LLMs. Because all representation variants are evaluated under the same model and decoding configuration, \emph{relative} comparisons across representations remain meaningful within this study, but cross-model generality is unknown. Future work should evaluate frontier models, open-source code models, and reasoning-oriented models to assess whether the representation ordering is stable.

\subsection{Single-Run Stochasticity}

Each testcase-variant pair is evaluated with a single inference query. Even at low temperature ($T{=}0.2$), sampling randomness may affect individual predictions. As a result, small performance differences between variants (e.g., 1--3 percentage points) should be interpreted cautiously. Future work should run multiple queries per testcase-variant pair to obtain confidence intervals and test statistical significance.

\subsection{Prompting Strategy}

The benchmark uses a CoT-style prompting strategy with dynamic few-shot examples. More sophisticated strategies---self-consistency decoding, graph-guided prompting, retrieval-augmented context selection---may interact differently with representation types and could change both absolute accuracy and the relative ordering of variants.

\subsection{Few-Shot Selection Confound}

Dynamic few-shot selection may vary across prompt sizes and representations. Because few-shot examples are removed when the prompt budget is tight, some representation variants may receive different in-context guidance than others. Specifically, source-plus-graph variants---which are the largest---are most likely to have few-shot examples omitted, which may independently hurt their performance. This confound is partially conflated with the context dilution effect and should be disentangled in future work through ablations that control few-shot content independently.

\subsection{Dataset Scope}

The standard corpus covers five CWE categories and 107 PrimeVul-derived testcases. The per-CWE distribution is imbalanced (CWE-190 accounts for more than half of all cases). Findings should be interpreted as evidence for the selected arithmetic and memory-lifecycle vulnerability classes in C/C++ rather than as universal conclusions across all vulnerability types.

\subsection{Graph Extraction Constraints}

Joern-based extraction depends on successful parsing and graph construction. ePDG generation additionally requires build feasibility and external artifact availability. Some valid testcases cannot be evaluated under all representation conditions. The extraction pipeline's limitations may introduce selection bias if testcases that fail extraction systematically differ from those that succeed.

\subsection{Prompt Serialization Bias}

Results depend not only on the graph type but on how the graph is serialized into text. Alternative node ordering strategies, edge annotation formats, summarization policies, and clipping policies could affect model behavior. A more complete study should systematically ablate graph encoding strategies independently of graph type.

\section{Related Work}

\subsection{LLM-Based Vulnerability Detection}

In the last few years,
LLMs have gained substantial momentum in vulnerability detection, complementing traditional approaches 
based on static/dynamic code analysis~\cite{kroening2014cbmc,emamdoost2021detecting,wen26fsetool,nong2021evaluating} and 
those based on deep-learning (DL)~\cite{nong2022open,mirskyvulchecker}, including 
those fine-tuning smaller pre-trained LLMs~\cite{nongvulgen,nongvgx,yang2024learning}. 
Inputs to both classes of approaches have been in the form of source/executable code and 
graph representations of programs (e.g., AST, PDG, and CFG)~\cite{shanmugasundaram2025deep}, 
including single forms and combined ones (e.g., AST+CFG+DFG~\cite{zhou2019devign}) like what we explore in this study.

More recent work has evaluated LLMs for vulnerability detection, localization, and repair across multiple programming languages~\cite{nong2024chain,zhou2024largelanguagemodelvulnerability}. Dominant techniques include fine-tuning on labeled vulnerability corpora, chain-of-thought prompt engineering, and retrieval-augmented prompting with specification or documentation context~\cite{nong2025appatch,guangbei26icse,li2025svtrustevalcevaluatingstructuresemantic,yu26tosem}. Several systems rely on raw source-code prompting or code-oriented fine-tuning with no explicit program structure. This paper complements these works by isolating representation as the primary experimental variable: rather than proposing a new detector, we empirically characterize how structural representation choices affect LLM reasoning quality for a fixed detection task.

\subsection{Graph-Based Vulnerability Analysis}

Graph-based vulnerability analysis has a rich history in both traditional static analysis and learning-based detection. Code Property Graphs (CPGs)~\cite{joern_paper}, which unify AST, CFG, PDG, and call-graph information into a single representation, have been used to query for vulnerability patterns and to train neural classifiers over program structure. Prior work on learning over PDGs, data-flow graphs, and code gadgets has demonstrated that explicit structural representations improve vulnerability detection over purely sequence-based approaches~\cite{code_gadgets_paper,liu2024pretrainingpredictingprogramdependencies}.

Most closely related is concurrent work on LLMxCPG~\cite{lekssays2025llmxcpgcontextawarevulnerabilitydetection}, which uses graph-guided context extraction for LLM-based vulnerability detection. Our study is distinct in that we (1) systematically ablate individual graph types and their combinations rather than using a fixed extraction strategy, (2) measure prompt-overhead tradeoffs across representation variants, and (3) include externally generated ePDG artifacts from the VulChecker/Hector pipeline to explore richer vulnerability-oriented dependence metadata. Together, these studies suggest that graph-based prompt construction is a productive design space for LLM security reasoning.

\subsection{Neural Code Representations}

Prior neural code models have explored token sequences, AST paths, data-flow graphs, CPGs, and learned graph embeddings for code understanding tasks~\cite{sheng2025llmssoftwaresecuritysurvey,zhou2024largelanguagemodelvulnerability}. Controlled studies of how representation choice affects code understanding---including vulnerability detection---have been conducted in the neural model setting~\cite{namavar2022controlledexperimentdifferentcode,swarna2023impactmultiplesourcecode}. 
For instance, in their latest relevant study~\cite{sunabstract26tosem}, Sun et al. systematically investigated 
how various ways of AST parsing, AST preprocessing, and AST encoding affect AST-based code representation learning for 
subsequent tasks including code clone detection, code search, and code summarization~\cite{sunabstract26tosem}. 
Our study is complementary: we focus specifically on LLM \emph{prompting} (rather than neural model training) and on the prompt-overhead tradeoffs that arise when structural representations are serialized into text.

\subsection{Prompt Engineering and Context Selection for Code}

Prompt design and context selection are increasingly important for LLM-based software engineering. Research has shown that context length, example selection, instruction framing, and domain knowledge significantly affect model performance on code tasks~\cite{liu2023lostmiddlelanguagemodels}. 
Similarly to our Source-plus-graph prompt family combining source code with AST/PDG in prompting LLMs, 
VerLog~\cite{guo2025verlog} and FreeLens~\cite{guo2026freelens} combine source code with call graphs (CG) to form graph-based 
LLM prompts. 
Unlike our study, however, 
they both use differences in those program representations (i.e., code diff and CG diff)
to reason about functional code-change semantics between two programs, as opposed to the single-program representations in our prompting for reasoning about vulnerability semantics. 

The ``lost in the middle'' phenomenon~\cite{liu2023lostmiddlelanguagemodels}, in which LLMs perform worse when relevant information appears in the middle of long prompts, is directly related to the context dilution effect we observe. Our results extend this line of research by demonstrating that the \emph{choice of representation}---not merely the volume of context---is a primary driver of LLM vulnerability reasoning quality.

\section{Future Directions}

The immediate next step is to expand the benchmark beyond the current study. First, the evaluation should include a substantially wider set of models, especially modern frontier models, open-source code models, and reasoning-oriented models, to assess whether the representation ordering observed here generalizes across architectures. Second, the dataset should be expanded beyond the PrimeVul-derived C/C++ corpus to encompass more CWE categories and additional languages such as Python, Java, JavaScript, and Rust. Third, the prompt-overhead analysis should be extended with per-variant inference latency and approximate dollar cost per 100 testcases to quantify the full deployment tradeoff between representation types.

A further direction is to build a representation-aware LLM vulnerability detector using the best-performing representation strategy identified here. Such a system should be evaluated against raw-source prompting baselines, existing LLM-based vulnerability detectors, and traditional static-analysis tools. This would move the work from a representation-ablation benchmark toward a deployable security analysis system.

Future work should also study graph encoding strategies more systematically. The current benchmark uses compact line-centric serialization with simplified dependency labels, but alternative encodings---adjacency-list formats, explicit edge-type annotation, path-based encodings, vulnerability-guided subgraph extraction---may further improve reasoning quality. In particular, the serialization of node identifiers as opaque cross-references rather than meaningful labels is a known limitation of the current format.

Finally, the few-shot selection confound identified in \Cref{sec:threats} warrants dedicated ablation: future experiments should evaluate fixed few-shot configurations alongside the dynamic selection policy to separate the contribution of in-context example quality from representation quality.

\section{Conclusion}

This paper presented \tool{}, an empirical benchmark for evaluating how static-analysis-based program representations affect LLM vulnerability reasoning. Using a PrimeVul-derived corpus of 107 real-world C/C++ testcases across five CWE categories, we systematically evaluated raw source code, ASTs, CFGs, PDGs, graph combinations, source-plus-graph hybrids, and an auxiliary ePDG track under a fixed LLM vulnerability detection task.

The results establish that representation choice has a substantial and consistent effect. Focused graph-based prompts outperform raw source code by large margins: AST+PDG achieves 83.2\% curated accuracy versus 53.5\% for the raw-source baseline. Graph-only representations also provide the best accuracy--prompt-overhead tradeoff: source-plus-graph variants use 2.4$\times$ as many characters and 2.2$\times$ as many input tokens while underperforming graph-only prompts by 5 percentage points. Most surprisingly, source-plus-graph prompts do not improve over graph-only prompts and often underperform them, revealing a \emph{context dilution} effect that operates even without prompt truncation.

These findings call for a re-examination of current practice in LLM-based software security. Systems that rely solely on raw source-code prompting may miss decades of accumulated progress in static program analysis. Instead, future systems should use static analysis as a prompt-construction layer---extracting compact, security-relevant structural evidence and presenting it to LLMs in focused, representation-aware prompts. This paper provides the initial empirical foundation for such an approach.

\section*{Artifact and Reproducibility}

The project artifact has been open-sourced~\cite{repbench_artifact}. Its README documents scripts for Joern/ePDG graph extraction, graph serialization, prompt generation, model evaluation, and paper-plot generation, together with compact prompt templates/examples and retained result files under \path{primevul_paper_results_latest}. The artifact also includes the token-counting script, per-prompt token JSONL records (including exact/proxy reconstruction status), and summary data used for \Cref{tab:overhead}. Testcase identifiers in the case studies and result records support tracing reported examples back to their project, CWE, expected label, and representation outcome.

\bibliographystyle{plain}
\bibliography{references}

\end{document}